Cavendish–HEP–94/15
September 1994

# QCD and Jet Physics*

B.R. Webber

Cavendish Laboratory, University of Cambridge
Madingley Road, Cambridge CB3 0HE, U.K.

**Abstract**

The current status of the QCD coupling constant $\alpha_S$ and experimental and theoretical studies of hadronic jets are reviewed.

## 1. Introduction

The dynamics of strong interactions at high momentum scales, corresponding to short distances, remains a highly active field of investigation. This is the domain in which perturbative quantum chromodynamics is expected to give its most reliable predictions. At the same time, non-perturbative phenomena can never be entirely forgotten; after all, we observe jets of hadrons in our detectors instead of the quarks and gluons (partons) with which perturbative QCD is concerned.

The connection between jets and partons has been clarified considerably over recent years: this will be one of the the main topics of the present review. There has been remarkable progress on the experimental side, due to several factors: the high statistics obtained by the LEP and Tevatron experiments; the improvements in detectors, especially in secondary vertex detection (important for heavy quark jet tagging) and particle identification; new tools for data analysis, especially new jet algorithms and improved QCD Monte Carlo programs; and of course the advent of HERA, which promises to be a powerful instrument for studies of QCD and jet physics.

On the theoretical side, progress has been slower. The calculation of higher order corrections to the $e^+e^-$ hadronic cross section and to the $Z^0$ boson and $\tau$ lepton hadronic widths [1] continues. At this meeting, quartic mass corrections up to second order were reported [2, 3] and methods for estimating uncalculated higher-order corrections were presented [4]. Calculations of loop contributions to multi-jet production amplitudes have been performed using powerful new techniques based on string theory [5] and the helicity method [6]. General programs for the computation of higher-order jet cross sections in hadron-hadron collisions have been developed and used for detailed comparisons with experiment [7]. Work on the next-to-leading corrections to the cross section for $e^+e^- \to$ 4 jets is under way but not yet completed. Resummation of logarithmically enhanced terms to all orders has been carried out for $e^+e^-$ jet rates [8] and for several event shape variables to next-to-leading logarithmic accuracy [9]. This has allowed $\alpha_S$ determinations beyond the range of applicability of fixed-order predictions [10].

Much interesting work on QCD, both theoretical and experimental, has also been done recently in areas other than jet physics, notably on structure functions, heavy quark systems and lattice gauge theory. Because there were separate sessions devoted to these topics, they are dealt with here only in so far as they have been used to determine the strong coupling constant $\alpha_S$. The reader is referred to the relevant plenary talks [11, 12, 13] and parallel sessions for further coverage.

After a review of the status and recent measurements of $\alpha_S$ in sect. 2, the present report will focus in sect. 3 on the dynamics of jet fragmentation, concentrating in sect. 4 on comparative studies of jets from different sources. Other topics, discussed in sect. 5, include tests of the fundamental colour structure of the theory, direct photon production in hadron collisions, and the new subject of jet production in diffractive processes. Some conclusions are presented in sect. 6.

## 2. The strong coupling constant

A wide range of methods are available for measuring the strong coupling $\alpha_S$, and many new measurements have been reported at this meeting or recently in other places. Before reviewing them, we should recall briefly the rather peculiar properties of this "fundamental constant".

### 2.1. Aspects of $\alpha_S$

First of all, $\alpha_S$ is not a constant but a running quantity, whose value depends on the energy scale $Q$ according to the equation

$$Q\frac{\partial\alpha_S}{\partial Q} = -\beta(\alpha_S) = -\frac{\beta_0}{2\pi}\alpha_S^2 - \frac{\beta_1}{8\pi^2}\alpha_S^3 - \cdots \quad (1)$$

where $\beta_0 = 11 - \frac{2}{3}N_f$ and $\beta_1 = 102 - \frac{38}{3}N_f$ for $N_f$ quark flavours. To this two-loop accuracy, the solution is†

$$\alpha_S(Q) = \frac{4\pi}{\beta_0 L}\left[1 - \frac{\beta_1 \ln L}{\beta_0^2 L} + \cdots\right] \quad (2)$$

where $L = \ln(Q^2/\Lambda^2)$, $\Lambda$ being a fundamental scale. All measurements of $\alpha_S(Q)$ at different values of $Q$ should correspond to a common value of $\Lambda$, or equivalently to a common value at some particular scale, usually taken nowadays to be the $Z^0$ boson mass $M_Z$.

In fact, $\alpha_S$ is not itself directly observable but is an auxiliary quantity in terms of which we may represent the perturbative prediction of any dimensionless quantity in the form

$$A(Q) = A_0 + A_1\alpha_S(\mu) + A_2(Q/\mu)\alpha_S^2(\mu) + \cdots \, . \quad (3)$$

If we could compute to all orders, the dependence on the arbitrary *renormalization scale* $\mu$ would cancel completely between $\alpha_S$ and the coefficients $A_i$. But if we choose $\mu$ very different from the natural scale $Q$ then large logarithms of $Q/\mu$ remain uncancelled in any finite order, making the prediction unreliable. In that sense we may say that an experiment at scale $Q$ is really measuring $\alpha_S(Q)$.

Another point to remember is that $\alpha_S$, not being directly observable, depends in higher order not only on the scale but also on the renormalization scheme. The customary reference scheme is nowadays the modified minimal subtraction ($\overline{\rm MS}$) scheme [15], so that the quoted scale $\Lambda$ is usually $\Lambda^{(5)}_{\overline{\rm MS}}$, the $\overline{\rm MS}$ scale with 5 active quark flavours, as appropriate at $Q = M_Z$. Note however that this scale (whose value is about 200 MeV) is not a very physical quantity: it is the scale at which the 5-flavour formula (2) would diverge if extrapolated far outside its domain of validity. Consequently it has become more usual to interpret all measurements in terms of a corresponding value of $\alpha_S(M_Z)$, using the three-loop version of eq.(2), with appropriate matching of different numbers of active flavours [16], to evolve from scale $Q$ to $M_Z$, which is what we shall do here.

Finally, before discussing particular measurements of $\alpha_S$, I should comment on the phenomenon of the "incredible shrinking error". Since $\alpha_S(Q)$ is a function of $Q/\Lambda$, eq. (1) tells us that the error in $\Lambda$ is related to that in $\alpha_S$ at any scale by

$$\frac{\delta\Lambda}{\Lambda} = \frac{\delta\alpha_S}{\beta(\alpha_S)} \, . \quad (4)$$

Since the $\beta$-function decreases with increasing scale like $\alpha_S^2$, this means that the best relative precision is obtained at the lowest possible scale, where $\alpha_S$ is largest. In terms of $\alpha_S(M_Z)$, we have

$$\frac{\delta\alpha_S(M_Z)}{\alpha_S(M_Z)} \sim \frac{\alpha_S(M_Z)}{\alpha_S(Q)}\frac{\delta\alpha_S(Q)}{\alpha_S(Q)} \, . \quad (5)$$

Thus the relative error in $\alpha_S$ is shrunk by the ratio of $\alpha_S(M_Z)$ to $\alpha_S(Q)$. However, we must bear in mind that perturbative and non-perturbative corrections that may be negligible at high scales can become important at $Q \ll M_Z$, so the net gain from measuring at a low scale is not so clear. In particular we need to look very carefully at possible sources of power-suppressed corrections ($1/Q^p$), which form a topic I shall return to repeatedly in later sections.

### 2.2. Status of $\alpha_S$ before this meeting

Figure 1 shows a summary of $\alpha_S$ values, averaged over various classes of measurement methods, prepared earlier by S. Bethke [17]. The values of $\alpha_S$ are given at the typical energy scales $Q$ at which the measurements were performed.

At the lowest scale comes the measurement based on the Gross-Llewellyn Smith (GLS) sum rule for deep inelastic neutrino-nucleon scattering. The perturbative corrections to the GLS sum rule are known up to $\mathcal{O}(\alpha_S^3)$ (i.e. next-to-next-to-leading order, NLLO) and the power ("higher twist") corrections to deep inelastic scattering are relatively well understood, making this an attractive method for determining $\alpha_S$ in spite of the low scale. The same comments apply to the other low-scale measurement shown, which uses the hadronic decays of the $\tau$ lepton. I discuss this method in detail below in connection with the new data presented at this conference.

† In fact $\alpha_S$ is known to three-loop accuracy [14].

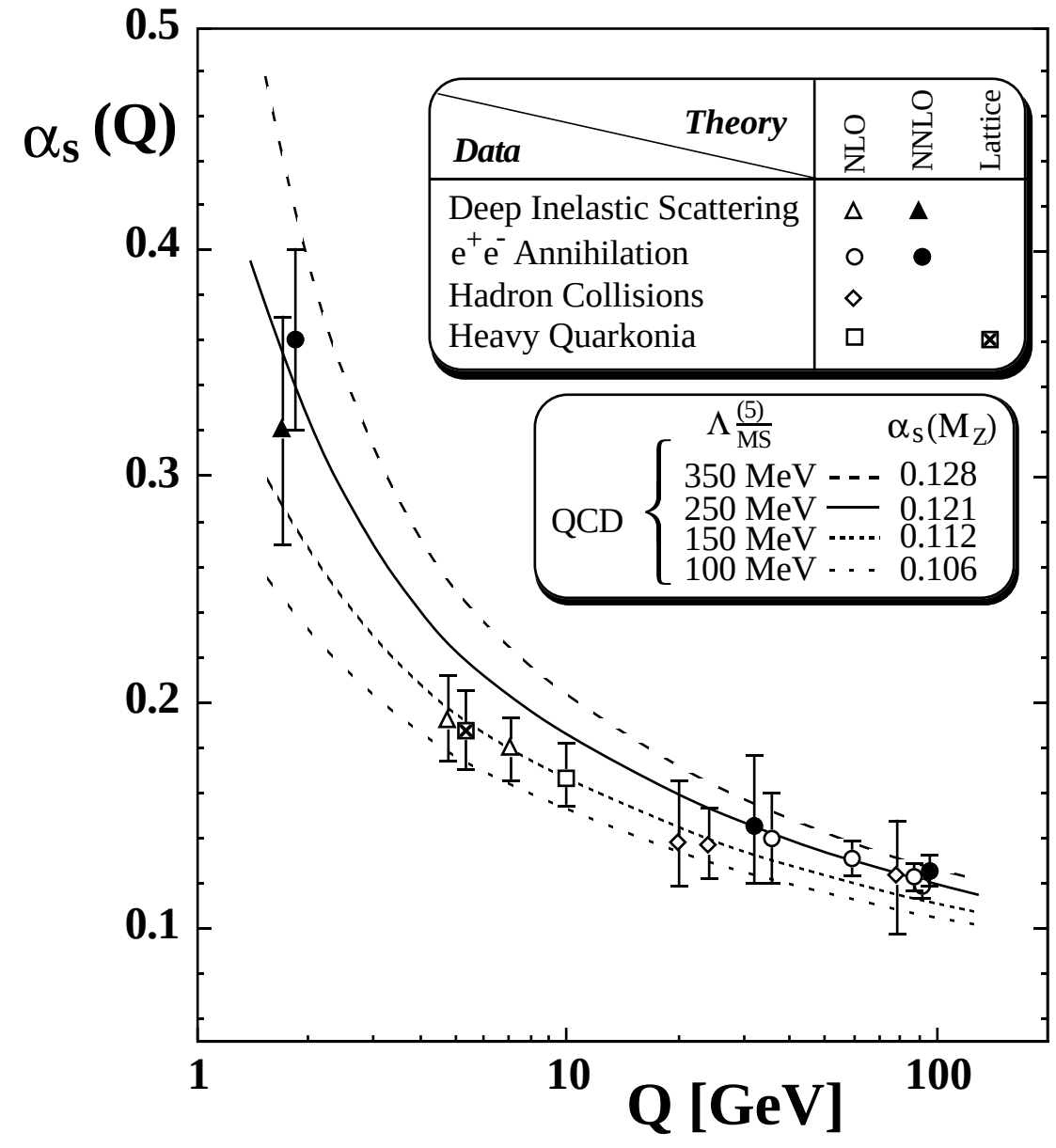

**Figure 1.** Summary of $\alpha_S$ measurements at various scales before this conference.

Moving up to intermediate scales, around $Q = 5$ GeV, the measurements based on the violation of Bjorken scaling in deep inelastic lepton-nucleon scattering have provided the most reliable information on $\alpha_S$. Here the calculations have been performed only to next-to-leading order (NLO), but again the higher-twist power corrections appear well under control.

A quite different method of measurement now coming into use involves the comparison of heavy quarkonium spectra with the non-perturbative predictions of lattice simulations of QCD. Here the relevant energy scale, set by the lattice spacing, is around 5 GeV for the current simulations. Again, this method is reviewed in more detail below in the context of new results.

At 10 GeV, the results shown are deduced from comparisons between data on $\Upsilon$ decays and perturbative predictions, some aspects of which we also discuss below.

Determinations of $\alpha_S$ at hadron-hadron (pp and $\mathrm{p\bar{p}}$) colliders, have been performed by comparing heavy quark, direct photon and W-boson plus jet cross sections with next-to-leading order predictions. Measurements based on pure jet production are not yet possible because the NLO corrections to multijet production have not been computed. So far, the precision achieved in hadron-hadron determinations has been less than that in $e^+e^-$ and lepton-hadron processes, owing to the larger experimental and theoretical uncertainties associated with incoming hadrons.

At high scales, in the range $30 - 100$ GeV, the majority of $\alpha_S$ determinations have come from $e^+e^-$ colliders, using either the total hadronic cross section (computed to NNLO) or NLO analyses of jet rates and event shapes. Latest results using these methods are also discussed below.

The results summarized in figure 1 are consistent with a 5-flavour $\overline{\mathrm{MS}}$ scale value of [17]

$$\Lambda^{(5)}_{\overline{\mathrm{MS}}} = 195^{+80}_{-60}\ \mathrm{MeV} \tag{6}$$

corresponding to

$$\alpha_S(M_Z) = 0.117 \pm 0.006\ . \tag{7}$$

One may, however, discern some possible systematic deviations from this value at intermediate scales in figure 1, which we shall discuss later in the light of the new data.

### *2.3. New results on $\alpha_S$*

While the compilation summarized above comprises essentially all the published information on $\alpha_S$, many new results have been presented at this conference or elsewhere in the last few months. Although most of them are preliminary and may therefore be subject to revision before publication, they form a valuable body of information which it seems appropriate to review critically at this time, always bearing in mind its provisional nature. This I do below, roughly in order of increasing energy scale. I then present a provisional update of averages for the various measurement methods, concluding with a global "preferred value" based on all information available at present.

*2.3.1. Bjorken sum rule.* Starting at the very lowest scales ($Q^2 \sim 2 - 3$ GeV$^2$), mention should be made of a recent theoretical study [18] in which the CERN EMC/NMC [19, 20] and SLAC E142/143 [21, 22] data on polarized lepton-nucleon scattering are used to extract a value of $\alpha_S$ from the higher-order corrections to the Bjorken sum rule for $g_1^p - g_1^n$. The advantage of this method is that the $\alpha_S$ dependence has been calculated up to $\mathcal{O}(\alpha_S^3)$ and there are estimates to $\mathcal{O}(\alpha_S^4)$. Non-perturbative corrections are expected to be $\mathcal{O}(1/Q^2)$ and there are arguments suggesting that the coefficient is small. The result obtained without such corrections‡

$$\begin{aligned}\alpha_S(Q^2 = 2.5\,\mathrm{GeV}^2) &= 0.375^{+0.062}_{-0.081}\\ \Rightarrow \alpha_S(M_Z) &= 0.122^{+0.005}_{-0.009}\end{aligned} \tag{8}$$

is certainly encouraging: one gets the full benefit of the "incredible shrinking error". What is needed to put this method on a firm footing is experimental evidence on the $Q^2$ dependence, to show convincingly that power corrections are indeed negligible even at these very low scales.

‡ The symbol '⇒' indicates a value obtained by evolving $\alpha_S$ from a lower scale.

*2.3.2. Hadronic $\tau$ decay.* One of the most carefully studied methods for measuring $\alpha_S$ is based on the hadronic decays of the $\tau$ lepton. Here not only the total hadronic width but also spectral moments of the hadron mass distribution have been calculated to $\mathcal{O}(\alpha_S^3)$ [23] and measured with high precision [24].

At the very low scale of these measurements, $Q \sim m_\tau$, one has to be sure that not only perturbative but also non-perturbative, power-suppressed corrections are well under control. Fortunately, a great deal can be inferred about power corrections from the operator product expansion (OPE) [23]. There should be terms of the form $c_D^i \langle O(D)\rangle / m_\tau^D$ where $D = 4, 6, \ldots$, $c_D^i$ is a computable coefficient for the $i$-th spectral moment, and the parameters $\langle O(D)\rangle$ describe vacuum condensates of the corresponding dimension [25]. There should be no $D = 2$ terms apart from those induced by the (current) quark masses, which are negligible except for decays to $s\bar{u}$. The $D = 4$ terms should correspond to the gluon condensate $\langle \alpha_S G^2 \rangle$ measured in other processes. All these predictions are well confirmed by the data [24, 26], and in particular there is no evidence of power corrections outside the framework of the OPE.

At this meeting a new preliminary result, based on a high-statistics analysis of the total width and a range of moments, was presented by the CLEO collaboration [27]:

$$\begin{aligned} \alpha_S(m_\tau) &= 0.309 \pm 0.024 \quad \text{(CLEO)} \\ \Rightarrow \alpha_S(M_Z) &= 0.114 \pm 0.003 \,. \end{aligned} \tag{9}$$

This new result agrees with the published ALEPH measurement [24] of $\alpha_S(m_\tau) = 0.330 \pm 0.046$. However, at the recent QCD94 meeting in Montpellier, the ALEPH collaboration presented a preliminary update of their result [28], based on increased statistics,

$$\begin{aligned} \alpha_S(m_\tau) &= 0.387 \pm 0.025 \quad \text{(ALEPH)} \\ \Rightarrow \alpha_S(M_Z) &= 0.124 \pm 0.003 \,. \end{aligned} \tag{10}$$

Thus there is currently some disagreement between the two experiments.

Looking at the data, one finds that the normalized spectral moments are in good agreement; the discrepancy lies entirely in the normalization. This is set by the leptonic branching fraction $B_l$ via the relation

$$\begin{aligned} R_\tau &\equiv \frac{\Gamma(\tau \to \nu_\tau \,\text{hadrons})}{\Gamma(\tau \to \nu_\tau \, e \, \bar{\nu}_e)} \\ &= \frac{1 - B_e - B_\mu}{B_l} = \frac{1}{B_l} - 1 - f_\mu \end{aligned} \tag{11}$$

where $f_\mu = 0.9726$ is a phase space correction. $B_l$ can be measured in three independent ways: from the $e$ and $\mu$ leptonic branching ratios, $B_e = B_l$ and $B_\mu = f_\mu B_l$, and from the lifetime $\tau_\tau = B_l \tau_\mu (m_\mu/m_\tau)^5$. ALEPH find somewhat lower values for $B_e$ and $B_\mu$, enhancing the hadronic rate, while their lifetime is consistent with that measured by CLEO, which is in turn consistent with the somewhat higher CLEO leptonic branching ratios.

Thus the current disagreement between the new high-precision results of the two experiments does not indicate any problem with the QCD analyses but turns entirely on the measured values of the leptonic branching ratios. We can expect that this issue will be resolved soon. For the moment, however, we take the mean of the two results (9) and (10), with an appropriately rescaled error, as a preliminary new measurement of $\alpha_S(M_Z)$ from $\tau$ decays:

$$\Rightarrow \alpha_S(M_Z) = 0.119 \pm 0.005 \,. \tag{12}$$

*2.3.3. Lattice QCD: $\alpha_S$ from $Q\bar{Q}$ spectra.* The lattice formulation of QCD remains the most promising approach that is independent of perturbation theory. So far, the most reliable $\alpha_S$ measurements using this method have come from comparing the computed spin-averaged quarkonium spectrum with data on the $c\bar{c}$ and $b\bar{b}$ systems. The treatment of heavy fermions has been handled either by the conventional Wilson formulation [29] or by the more recent technique of expansion around the non-relativistic limit (NRQCD) [30]. Generally the NRQCD approach has smaller quoted errors and leads to somewhat higher values of $\alpha_S$.

The previous results summarized in figure 1 were obtained in the quenched approximation, that is, neglecting the contributions of light quark loops, and the errors are dominated by the uncertainty in the corrections to this approximation. The error estimates are conservative in as much as they correspond roughly to the whole of the correction applied at the lattice scale.

Very recently, the first results of unquenched calculations including two flavours of dynamical quarks have been reported [31, 32]. The most precise is that using the NRQCD method and comparing with the $b\bar{b}$ spectrum [32], which gives

$$\alpha_S(5\ \text{GeV}) = 0.203 \pm 0.007 \tag{13}$$

$$\Rightarrow \alpha_S(M_Z) = 0.115 \pm 0.002 \,. \tag{14}$$

Here the scale set by the lattice spacing is in fact 8.2 GeV; the scale of 5 GeV quoted above is the closest at which a value is given after extrapolation to the physical number of flavours and conversion to the $\overline{\text{MS}}$ scheme. Since results are now available for $N_f = 0$ and 2, the estimated error in the extrapolation is small (0.2% in $\alpha_S(M_Z)$). The dominant source of systematic error is now stated to be the conversion from the static-quark potential (V) renormalization scheme to $\overline{\text{MS}}$ (1.7%), which could be reduced by a third-order perturbative calculation.

The unquenched calculation using Wilson heavy fermions compared with charmonium data [31] obtains consistent, although somewhat lower, results with larger errors.

*2.3.4. $\Upsilon$ decays.* A new determination of $\alpha_S$ from the ratio of radiative and non-radiative widths of the $\Upsilon$ were reported by the CLEO collaboration at the QCD94 meeting [33]. The idea is to use the ratio

$$\begin{aligned} \frac{\Gamma(\Upsilon\to\gamma+\text{had.})}{\Gamma(\Upsilon\to\text{hadrons})} &\sim \frac{\Gamma(\Upsilon\to\gamma gg)}{\Gamma(\Upsilon\to ggg)} \\ &\sim \frac{4\alpha_{\rm em}}{5\alpha_S(m_b)}\left[1-2.6\frac{\alpha_S}{\pi}\right] , \end{aligned} \qquad (15)$$

for which the quarkonium wavefunction cancels and relativistic corrections are likely to be minimized. The preliminary result is

$$\begin{aligned} \alpha_S(M_\Upsilon) &= 0.164\pm 0.003\pm 0.008\pm 0.013 \\ \Rightarrow \alpha_S(M_Z) &= 0.111\pm 0.006 \end{aligned} \qquad (16)$$

where the first error is statistical, the second experimental systematic, and the third theoretical.

It has only been appreciated quite recently that the process $\Upsilon \to \gamma$ + hadrons has an additional *leading-order* QCD correction due to the fragmentation of gluons into photons [34]. This is essentially because the specification of the final state is not fully inclusive: a real photon must be present. In higher order, quark fragmentation also contributes. The relevant fragmentation functions have to be determined experimentally. They probably have only a small effect, but this has yet to be checked in detail. Meanwhile, a somewhat larger estimate of the theoretical uncertainty (say $\Delta\alpha_S(M_Z) = 0.01$) may be advisable.

*2.3.5. $e^+e^-$ jet rates and event shapes.* The CLEO collaboration have also reported preliminary results [35] on $e^+e^-$ jet rates in the four-flavour continuum at c.m. energy $\sqrt{s} = 10.53$ GeV. In an analysis very similar to those at higher energies, they study the dependence of the two-jet rate (defined using the Durham or $k_\perp$ clustering algorithm [36]) on the resolution parameter $y_{\rm cut}$, to obtain

$$\begin{aligned} \alpha_S(10.53\,\text{GeV}) &= 0.164\pm 0.004\pm 0.014 \\ \Rightarrow \alpha_S(M_Z) &= 0.113\pm 0.007 , \end{aligned} \qquad (17)$$

where the first error is experimental and the second is theoretical. The latter is assessed to be dominated by the renormalization scale dependence of the $\mathcal{O}(\alpha_S^2)$ prediction.

At these relatively low energies there are substantial hadronization corrections to jet rates, which are estimated using the QCD Monte Carlo programs JETSET [37] and HERWIG [38]. After tuning both programs to the CLEO data at the hadron level, the difference between their predictions at the parton level is used to estimate the error due to hadronization. In this case the two programs agree well at the parton level and so the estimated hadronization error is small, $\Delta\alpha_S(M_Z) = 0.003$.

The widely-adopted prescription of assessing hadronization errors by comparing JETSET and HERWIG could be misleading, especially when the hadronization correction itself is large. The term 'parton level' in these programs means something different from that used in purely perturbative calculations, since the programs use different variables, cutoffs and approximation schemes for the parton shower phase of final-state development. Thus one is comparing inequivalent quantities. Furthermore, the fact that the programs have to be re-tuned at this energy, relative to higher energies, indicates that they do not reproduce fully the energy dependence of the corrections. This is an indication that the physics of hadronization is not handled perfectly by the programs, even if the re-tuned fits at this energy are very good.

A safer way to assess the hadronization error might be to base it on an agreed fraction (say $\pm 50\%$) of the *total* hadronization correction or the discrepancy between the JETSET and HERWIG corrections, whichever is the greater. This would favour observables for which the predicted corrections are small. Of course, this suggestion applies not only to the CLEO measurement but also to those at higher energies. However the assessed error is likely to be enlarged more at lower energies, owing to the general $1/Q$ energy dependence of hadronization effects (see below).

It should be emphasised at this point that it would be very worthwhile for CLEO to measure not only the jet rates but also, if possible, the total hadronic rate $R$ in the four-flavour continuum. This quantity would be less subject to hadronization uncertainties, and the calculations of higher-order and quark-mass corrections [2] are now of such precision that a very good $\alpha_S$ determination should be possible, especially taking into account the benefit to be obtained at these energies from the 'incredible shrinking error'.

Moving to higher energies, new measurements based on the 2-jet rate, this time with logarithms of $y_{\rm cut}$ resummed to all orders, were presented by the TPC/2$\gamma$ [39] and TOPAZ [40] collaborations at 29 and 58 GeV respectively. Their results are

$$\begin{aligned} \alpha_S(29\,\text{GeV}) &= 0.160\pm 0.012 \\ \alpha_S(58\,\text{GeV}) &= 0.139\pm 0.008 . \end{aligned} \qquad (18)$$

The interesting feature of these measurements is not so much their precision as the fact that the analysis in each case follows very closely that performed earlier at $\sqrt{s} = M_Z$ by the ALEPH collaboration [41]. Thus one has three essentially identical measurements at different scales, which show the running of $\alpha_S$ convincingly (figure 2).

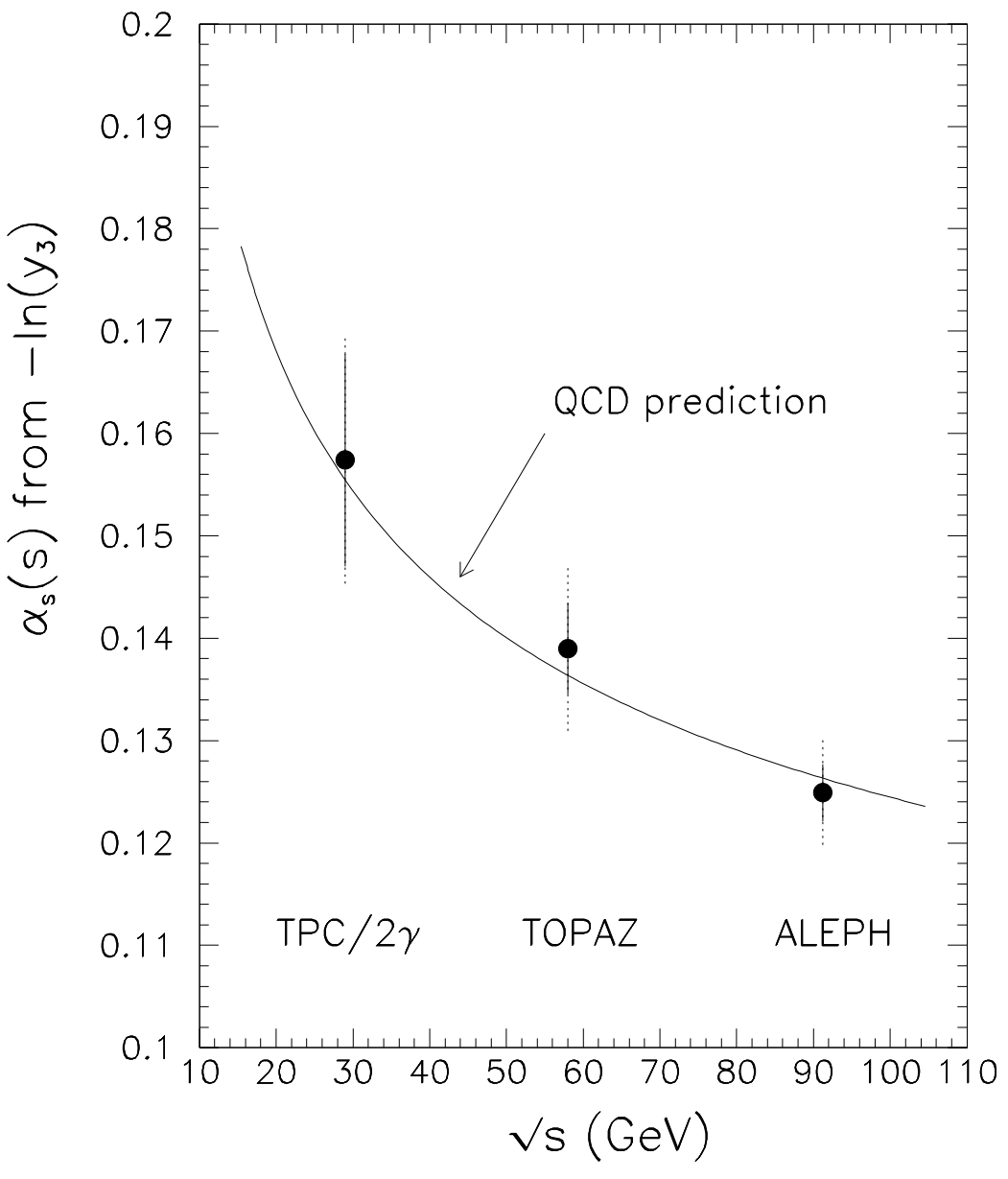

**Figure 2.** Running of $\alpha_S$ as seen from identical analyses of $e^+e^-$ jet rates at different energies.

New studies of hadronic final states in $Z^0$ decays reported at this meeting include an analysis of event shapes and jet rates by the SLD collaboration [42] , who find

$$\alpha_S(M_Z) = 0.120 \pm 0.003 \pm 0.009 \qquad (19)$$

from comparisons with resummed predictions. A novel analysis by the OPAL collaboration [43], using a cone jet algorithm in contrast to the jet clustering approach usually adopted in $e^+e^-$ studies, gives

$$\alpha_S(M_Z) = 0.119 \pm 0.008 \qquad (20)$$

from the dependence of jet rates on the minimum jet energy, and

$$\alpha_S(M_Z) = 0.116 \pm 0.008 \qquad (21)$$

from the dependence on the angular size of the cone, comparing in each case with $\mathcal{O}(\alpha_S^2)$ predictions.

A disadvantage of the cone jet algorithm is that predictions with resummation of large terms beyond $\mathcal{O}(\alpha_S^2)$ are not available. This is related to the fact that the treatment of multijet final states is more complicated, and more ambiguous, than for a clustering algorithm. One big advantage, however, is that results can be compared directly with those already obtained using a cone algorithm in hadron-hadron collisions. This will be discussed in sect. 4.2.

*2.3.6. Scaling violation in $e^+e^- \to hX$.* The violation of scaling in jet fragmentation seems a natural phenomenon to use for measuring $\alpha_S$, since it is closely analogous to scaling violation in deep inelastic structure functions, which has been used for many years for that purpose. As we shall discuss in sect. 3.1, there are many additional complications in the fragmentation measurement, and so it is only recently that results of this approach have become available.

The first analysis, by the DELPHI collaboration [44], relied heavily on the JETSET Monte Carlo to obtain a result without direct experimental input on the fragmentation of different quark flavours and gluons. The result was

$$\alpha_S(M_Z) = 0.118 \pm 0.005\,. \qquad (22)$$

One could assign an additional theoretical uncertainty to this result, with a magnitude depending on the level of commitment to JETSET. In addition, the DELPHI analysis used only the $\mathcal{O}(\alpha_S^2)$ matrix elements instead of the full machinery of the next-to-leading evolution equations for the fragmentation functions.

A new analysis of scaling violation by the ALEPH collaboration [45] avoids model dependence as far as possible and obtains the result

$$\alpha_S(M_Z) = 0.127 \pm 0.011 \qquad (23)$$

from a comparison with full next-to-leading evolution. The relatively large error reflects primarily the independence from model assumptions. This analysis and related issues will be discussed in some detail in sect. 3.

*2.3.7. $ep \to 2+1$ jet rate.* The advent of HERA raises the exciting prospect of measuring $\alpha_S$ in a variety of ways, over a wide range of $Q^2$, in a single experiment. At this meeting, preliminary results were reported [46] from a study by the H1 collaboration [47] of the $2+1$ jet rate, that is, the fraction of deep inelastic events in which one can resolve two final-state "current" jets in addition to the "beam" jet which carries the remnants of the incoming proton. The $2+1$ jet cross section is at least first order in $\alpha_S$ and $\mathcal{O}(\alpha_S^2)$ calculations, using a modified JADE jet clustering algorithm [48], are now available [49, 50]. The results on the rate and the extracted values of $\alpha_S$ in the range $Q^2 = 10 - 4000\ \mathrm{GeV}^2$ are shown in figure 3.

The jet rate for a given value of $\alpha_S$ depends on the parton distributions in the proton, and so one has to assume a set of distributions in order to deduce a value of $\alpha_S$. Results for two different MRS [51] sets of parametrizations are shown. Of these, the D0 set is now definitely ruled out by the HERA low-$x$ structure function data; the D– set is now reckoned to be a little too high at small $x$ but not so bad. A fuller discussion may be found in the relevant plenary talk [11]. For the present we may take the $\alpha_S$ values extracted using the D– set (figure 3d) as sufficiently reliable. They show clearly the power of the HERA experiments to observe the running of $\alpha_S$.

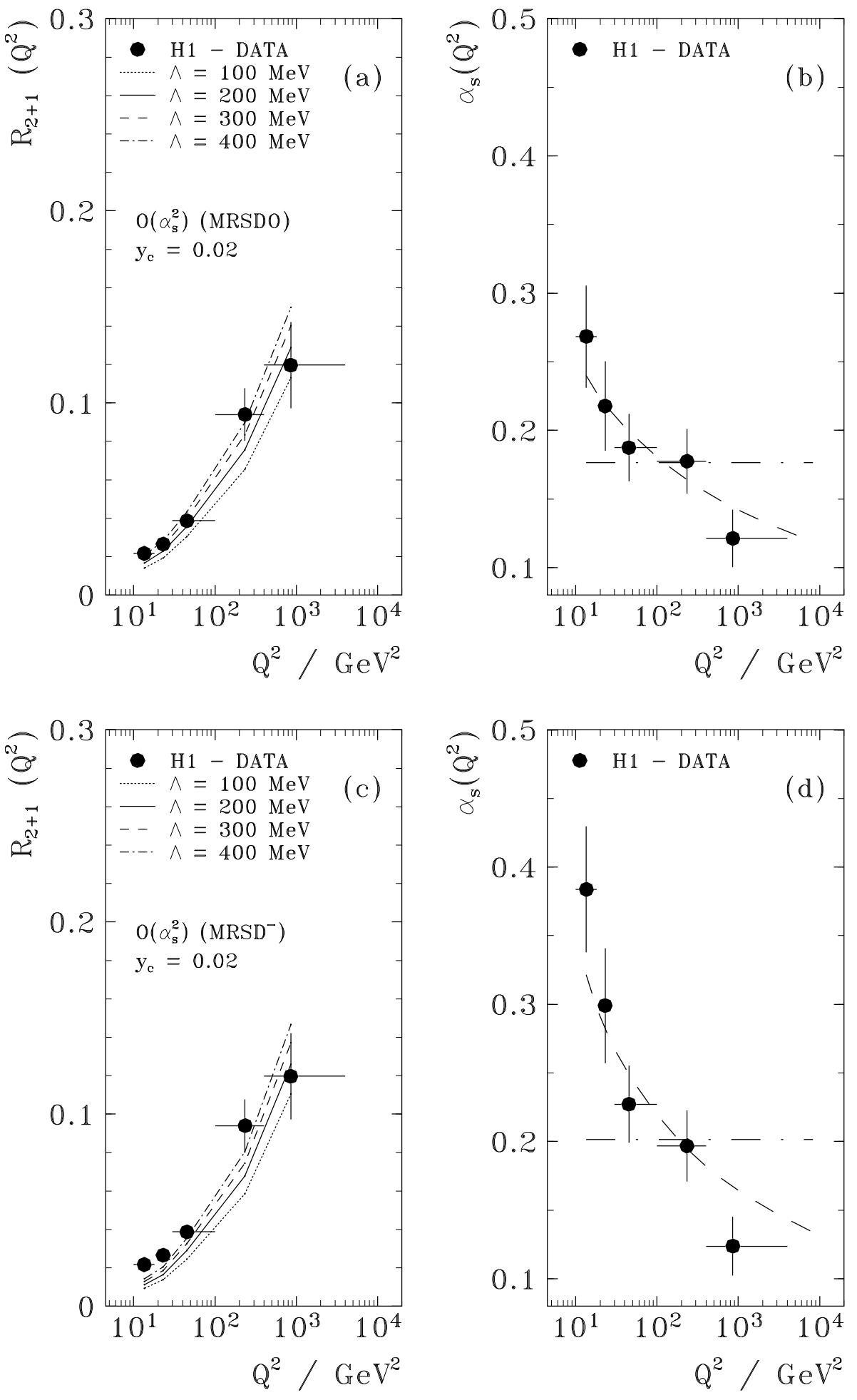

**Figure 3.** H1 results on the $2+1$-jet rate and $\alpha_S$ as functions of $Q^2$: (a,b) using MRSD0 parton distributions to extract $\alpha_S$; (c,d) using MRSD– parton distributions.

One can reduce the sensitivity to the choice of parton distributions by cutting out the region of low momentum fractions of the struck parton (called $x_p$, which is larger than the Bjorken variable $x$ for $2+1$ jet production). For $x_p > 0.01$ the difference between MRS D0 and D– is negligible. This cut eliminates the two lowest $Q^2$ points ($Q^2 < 30$ GeV$^2$) and extrapolation to $Q = M_Z$ yields the preliminary result

$$\Rightarrow \alpha_S(M_Z) = 0.121 \pm 0.015\ . \qquad (24)$$

We may expect the result (24), and corresponding results from the ZEUS experiment, to be refined considerably as statistics accumulate. For example, one will be able to fit the joint dependence on $Q^2$ and the jet resolution $y_{\rm cut}$, instead of fixing $y_{\rm cut} = 0.02$ as in figure 3. Figure 4 shows the ZEUS data [53] on the $y_{\rm cut}$ dependence of jet rates integrated over the region $160 < Q^2 < 1280$ GeV$^2$, $0.01 < x < 0.1$.

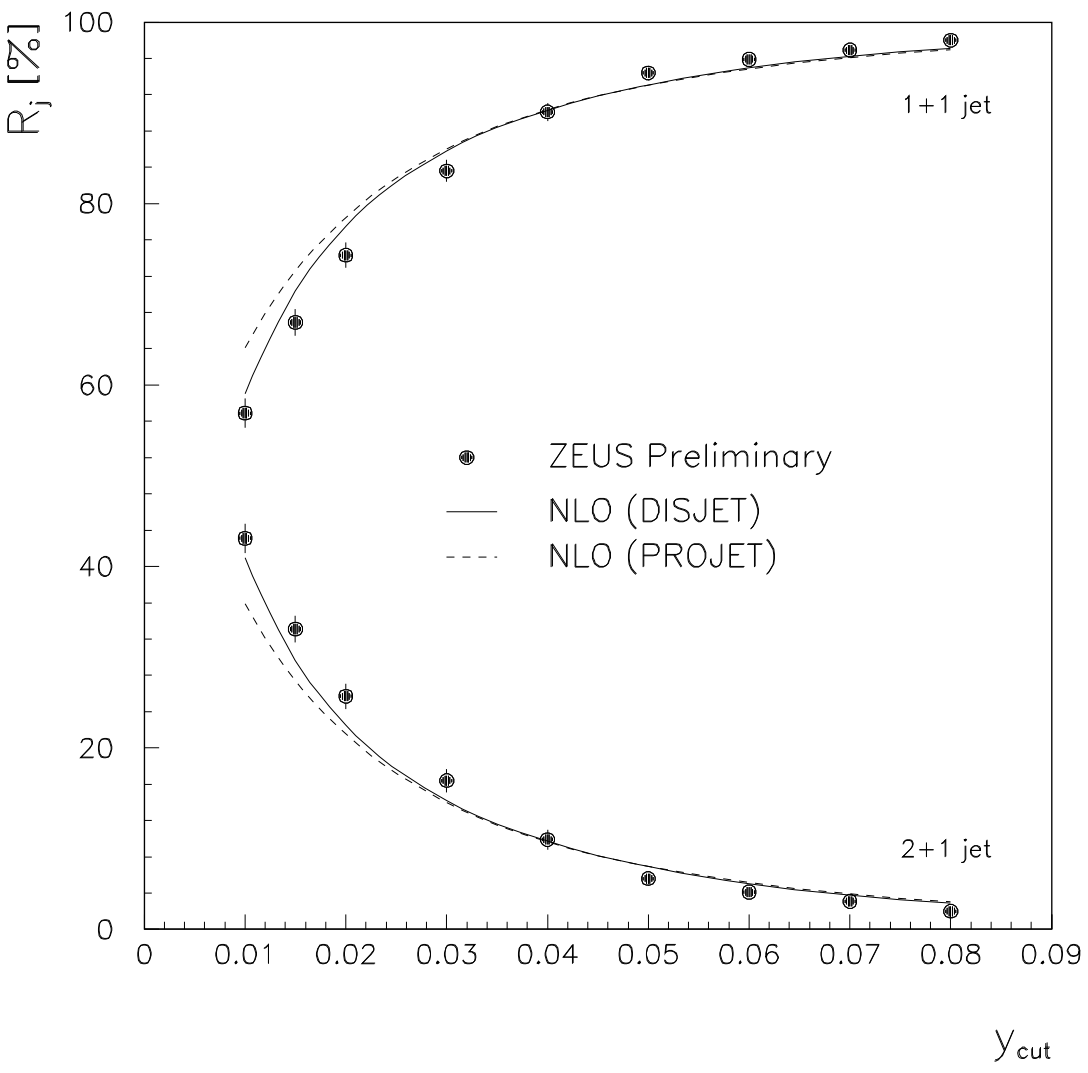

**Figure 4.** Results from ZEUS on the resolution scale dependence of ep jet rates.

The curves do not represent a fit but simply correspond to the $\mathcal{O}(\alpha_S^2)$ predictions for $\alpha_S(M_Z) = 0.124$. Clearly the fitted value would not be far from this.

As the HERA measurements are refined, one will need to pay close attention to the consistency between the output value of $\alpha_S$ and the value of $\Lambda_{\overline{\rm MS}}$ that is fed into the evolution of the assumed parton distributions, which is normally quite low (corresponding to $\alpha_S(M_Z) \simeq 0.113$ for MRS [51]). For strict consistency, one would need to re-evolve the distributions over the relevant region of $Q^2$ using the corresponding value of $\Lambda_{\overline{\rm MS}}$. A simpler alternative would be to rescale the values of $Q$ at which the distributions are evaluated, changing $Q$ to $Q\Lambda'/\Lambda$ when replacing $\Lambda$ by $\Lambda'$ in the parametrizations [52]. Since the QCD evolution depends only on the ratio $Q/\Lambda$, the scaling violation in the parton distributions would then be consistent with the new value $\Lambda_{\overline{\rm MS}} = \Lambda'$.

*2.3.8. $Z^0$ hadronic width.* Finally I should report the latest $\alpha_S$ value from the combined fits of the LEP Electroweak Working Group [54],

$$\alpha_S(M_Z) = 0.125 \pm 0.005 \pm 0.002\ . \qquad (25)$$

This number, determined primarily by the LEP data on the hadronic width of the $Z^0$, now represents one of the highest measured values, whilst remaining the most firmly-based from the theoretical point of view.

### *2.4. Summary of results on $\alpha_S$*

The new results discussed at this meeting are summarized in figure 5. It is impressive that these measurements alone cover the whole accessible energy range and show the running of $\alpha_S$ convincingly.

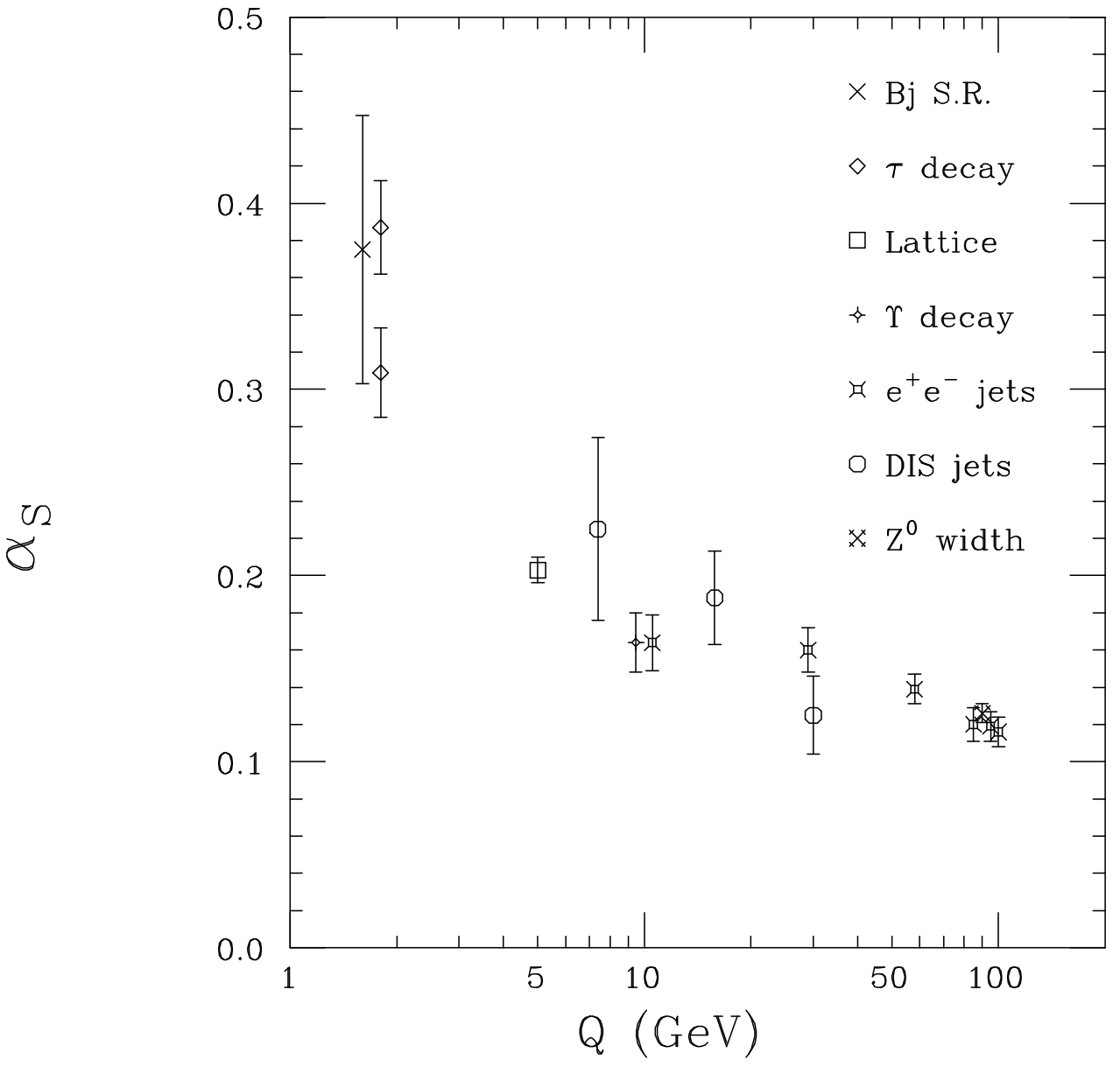

**Figure 5.** Summary of preliminary $\alpha_S$ measurements discussed here.

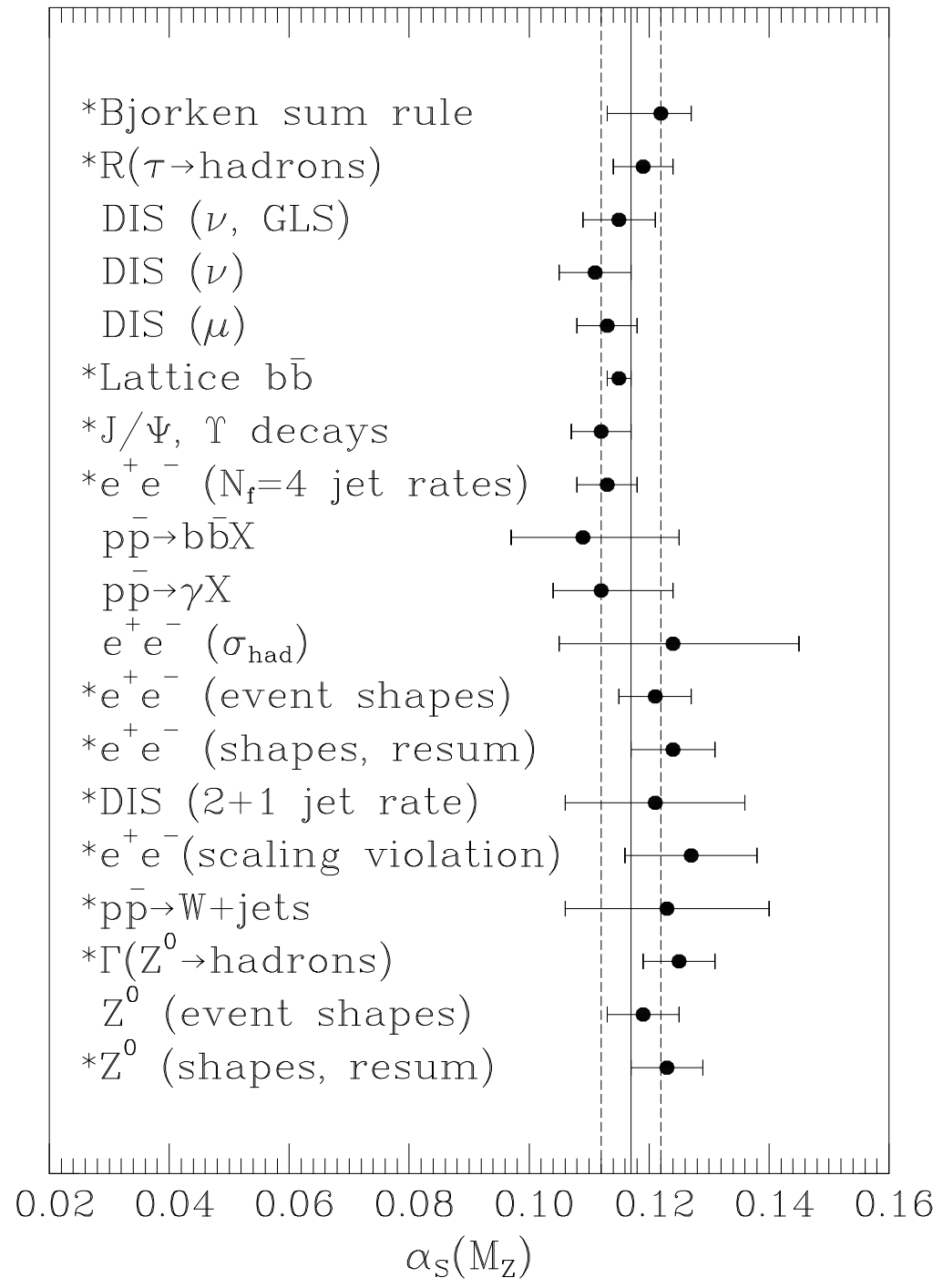

**Figure 6.** Summary of $\alpha_S$ measurements, evolved to the scale $M_Z$. Starred items include preliminary results discussed here.

Combining the new (albeit preliminary) results with the earlier ones and evolving all of them to $Q = M_Z$, we obtain figure 6. The indicated errors usually have large systematic and theoretical components, so it is difficult to quantify the level of agreement between the different methods of measurement. Naively computing chi-squared gives a good fit ($\chi^2$/d.o.f. = 11/18) with a mean of $\alpha_S(M_Z) = 0.1165 \pm 0.0012$. A more cautious approach to the error would be to assign an uncertainty equal to that of a typical measurement by a reliable method. This leads to my "global preferred value"

$$\Rightarrow \alpha_S(M_Z) = 0.117 \pm 0.005\,. \tag{26}$$

Notice that the mean value does not shift significantly from the current world average (7) based on published data. The slight reduction in the error reflects not so much the precision of the new measurements as my own caution in the assessment of possible systematic errors.

The results in figure 6, which are presented roughly in increasing order of energy scale, tend to lie slightly below the mean value (26) at low scales (say, below $Q = 30$ GeV) and above it at higher scales. This tendency is not strong enough to take very seriously at present. Nevertheless it is worthwhile to consider possible causes of such an effect.

One obvious possibility is that the running of $\alpha_S$ is not as expected.§ This would be the case if new types of coloured particles were contributing to the QCD $\beta$-function. One candidate would be a light gluino. However, gluino masses in the appropriate range are disfavoured by other data [55, 56]. For example, studies of $e^+e^- \to 4$ jets have found no sign of the production of non-standard coloured particles, as we shall discuss in sect. 5.1.

Another possibility is that there is a systematic difference between hard processes in which the large momentum scale is spacelike, as in deep inelastic scattering, and those in which it is timelike, as in $e^+e^-$ annihilation [57]. Then the association of low $\alpha_S(M_Z)$ with low scales would not be significant, being merely a reflection of the fact that deep inelastic data predominate at lower scales. The data from HERA will play a crucial rôle in testing this hypothesis. Clarification of the results from $\tau$ decay will also be helpful. If the running of $\alpha_S$ is at fault, this process should yield a low value because of its low scale. In the other scenario, it would give a high value because it is timelike.

A third possible source of discrepancy that has been suggested [58] is quark mass effects. The lowest $\alpha_S$ values seem to be concentrated in the intermediate region of $Q$ in which heavy quarks start to play a rôle. Most higher-order QCD corrections have been calculated only in the massless quark limit, and might therefore be unreliable in this region. An exception is the total $e^+e^-$ hadronic rate [2], making a precision measurement of this quantity at an intermediate scale very desirable, as already mentioned in sect. 2.3.5.

§ It is for this reason that I have tried to distinguish between evolved and direct measurements of $\alpha_S(M_Z)$.

## 3. Jet fragmentation

### *3.1. Scaling violation in fragmentation*

One of the most basic features of a jet is the energy spectrum of the hadrons into which it fragments. As mentioned earlier, some recent $\alpha_S$ determinations have been based on the observation of scaling violation in fragmentation [44, 45]. One defines the total fragmentation function for hadrons of type $h$ in $e^+e^-$ annihilation at c.m. energy $\sqrt{s}$ as

$$F(x,s) = \frac{1}{\sigma_{\rm tot}}\frac{d\sigma}{dx}(e^+e^- \to hX) \qquad (27)$$

where $x = 2E_h/\sqrt{s}$. Normally $h$ represents any charged hadron and the approximation $x = x_p = 2p_h/\sqrt{s}$ is used.

The fragmentation function (27) can be represented as a sum of contributions from the different primary partons $i = u, d, \ldots, g$,

$$F(x,s) = \sum_i \int \frac{dz}{z} C_i(s;\alpha_S,z) D_i(x/z,s) \ . \qquad (28)$$

To zeroth order in $\alpha_S$, the coefficient function $C_g$ for gluons is zero, while for quarks $C_i = g_i(s)\delta(1-z)$ where $g_i(s)$ is the appropriate electroweak coupling. The parton fragmentation functions $D_i$ satisfy evolution equations of the Altarelli-Parisi type

$$Q^2 \frac{\partial}{\partial Q^2} D_i(x,Q^2) = \frac{\alpha_S}{2\pi} \sum_j \int \frac{dz}{z} P_{ij}(\alpha_S,z) D_i(x/z,Q^2) \ . \qquad (29)$$

The form of the splitting functions $P_{ij}$ is known to second order. Thus fitting the variation of $F(x,s)$ over a range of $s$ allows one to extract a value of $\alpha_S$.

There are several complications in the scaling violation analysis [59]. First, the energy dependence of the electroweak couplings $g_i(s)$ is especially strong in the energy region under study ($\sqrt{s} = 20 - 90$ GeV). In particular, the $b$-quark contribution increases markedly as one approaches the $Z^0$ mass (figure 7). The fragmentation of the $b$-quark into charged hadrons, including the decay products of the $b$-flavoured hadron, is expected to be substantially softer than that of the other quarks, so its increased contribution can give rise to a 'fake' scaling violation that has nothing to do with QCD. A smaller, partially compensating effect is expected in charm fragmentation. These effects can be eliminated by extracting the $b$- and $c$-fragmentation functions at $\sqrt{s} = M_Z$ from tagged heavy quark events, and evolving them separately to other energies. Unbiased results can be obtained by looking in the hemisphere opposite to a jet containing a tagging secondary vertex.

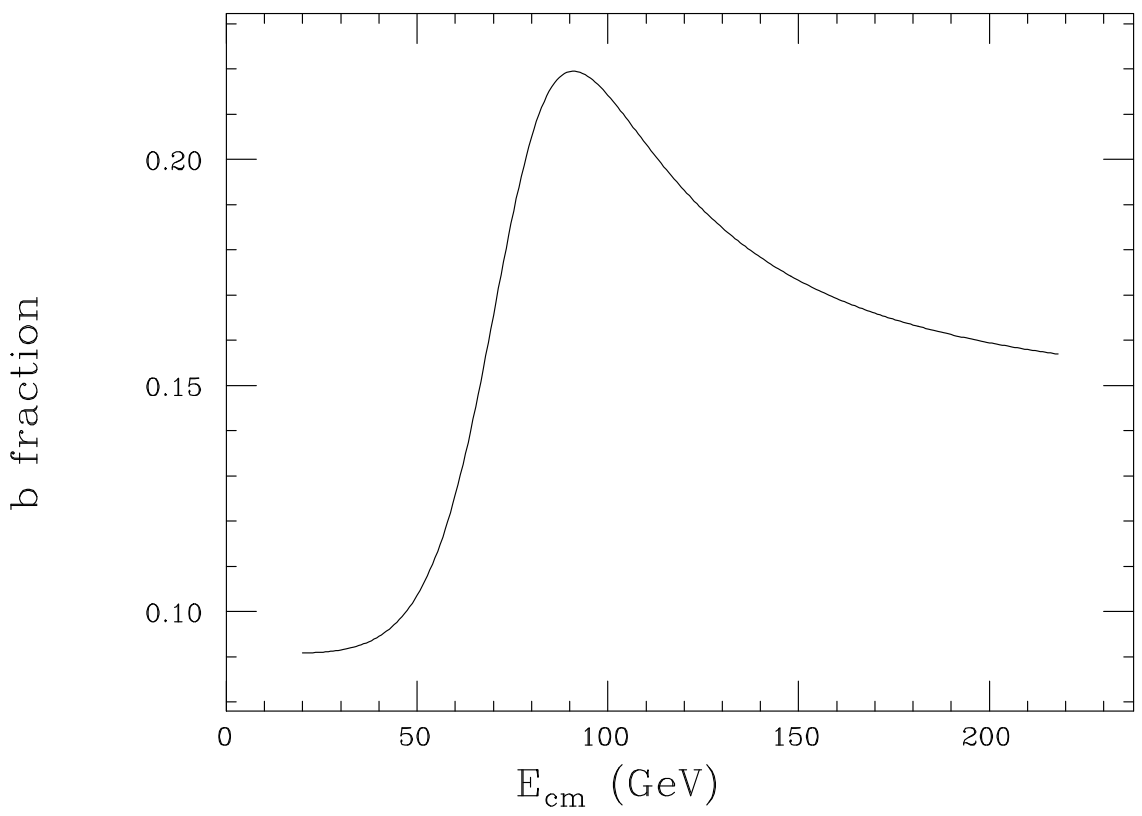

**Figure 7.** The $b$-quark fraction in $e^+e^- \to$ hadrons.

Secondly, one requires the gluon fragmentation function $D_g(x,s)$ in addition to those of the quarks. Although the gluon does not couple directly to the electroweak current, it contributes in higher order, and mixes with the quarks through evolution. Its fragmentation can be studied in tagged three-jet events, or via the longitudinal fragmentation function (see below).

A final complication is that power corrections to fragmentation functions, of the form $f(x)/Q^p$, are not well understood. As we shall discuss later, Monte Carlo studies suggest that hadronization can lead to $1/Q$ corrections. Therefore, possible contributions of this form should be included in the parametrization when fitting the scaling violation.

Preliminary results of an analysis along these lines by the ALEPH collaboration [45], based on a comparison of LEP data with those from lower-energy experiments, are shown in figure 8. Also included are their separate fragmentation functions for light $(u,d,s)$ quarks, $b$ quarks and gluons, showing that the latter two functions are significantly softer (quite similar to each other, within the errors) at $\sqrt{s} = M_Z$. An overall fit in the range $22 \leq \sqrt{s} \leq 91.2$ GeV, $0.1 < x < 0.8$, incorporating full next-to-leading-order evolution and a simple parametrization of power corrections, leads to the $\alpha_S$ value (23) above, and gives a good description of the data in the fitted region, as shown by the curves.|| The fitted power corrections are small and the value obtained for $\alpha_S$ is not highly sensitive to the form assumed for them.

### *3.2. Longitudinal fragmentation function*

At this meeting, the ALEPH [45] and OPAL [60] collaborations have presented preliminary results of the first analyses of the longitudinal fragmentation function in $e^+e^-$ annihilation. This is defined in terms of the joint distribution in the energy fraction $x$ and the angle $\theta$ between the observed hadron and the incoming electron beam [59]:

$$\frac{1}{\sigma_{\rm tot}}\frac{d^2\sigma}{dx\,d\cos\theta} = \frac{3}{8}(1+\cos^2\theta)\,F_{\rm T}(x) + \frac{3}{4}\sin^2\theta\,F_{\rm L}(x) + \frac{3}{4}\cos\theta\,F_{\rm A}(x)\;, \tag{30}$$

where $F_{\rm T}$, $F_{\rm L}$ and $F_{\rm A}$ are respectively the transverse, longitudinal and asymmetric fragmentation functions.¶ Like the total fragmentation function, $F = F_{\rm T} + F_{\rm L}$, each of these functions can be represented as a convolution of the parton fragmentation functions $D_i$ with appropriate coefficient functions $C_i^{\rm T,L,A}$ [61], as in eq. (28). In fact the transverse and longitudinal coefficient functions are related in such a way that

$$F_{\rm L}(x) = \frac{\alpha_{\rm S}}{2\pi}C_F\int\frac{dz}{z}\Big[F_{\rm T}(z) + 4\left(\frac{z}{x}-1\right)D_g(z)\Big] + \mathcal{O}(\alpha_{\rm S}^2)\;. \tag{31}$$

|| The fitted curves do not extrapolate well into the small-$x$ region, but a next-to-leading order treatment would not be expected to be reliable there, owing to the presence of large higher-order terms proportional to $\alpha_S^n \ln^{2n} x$.

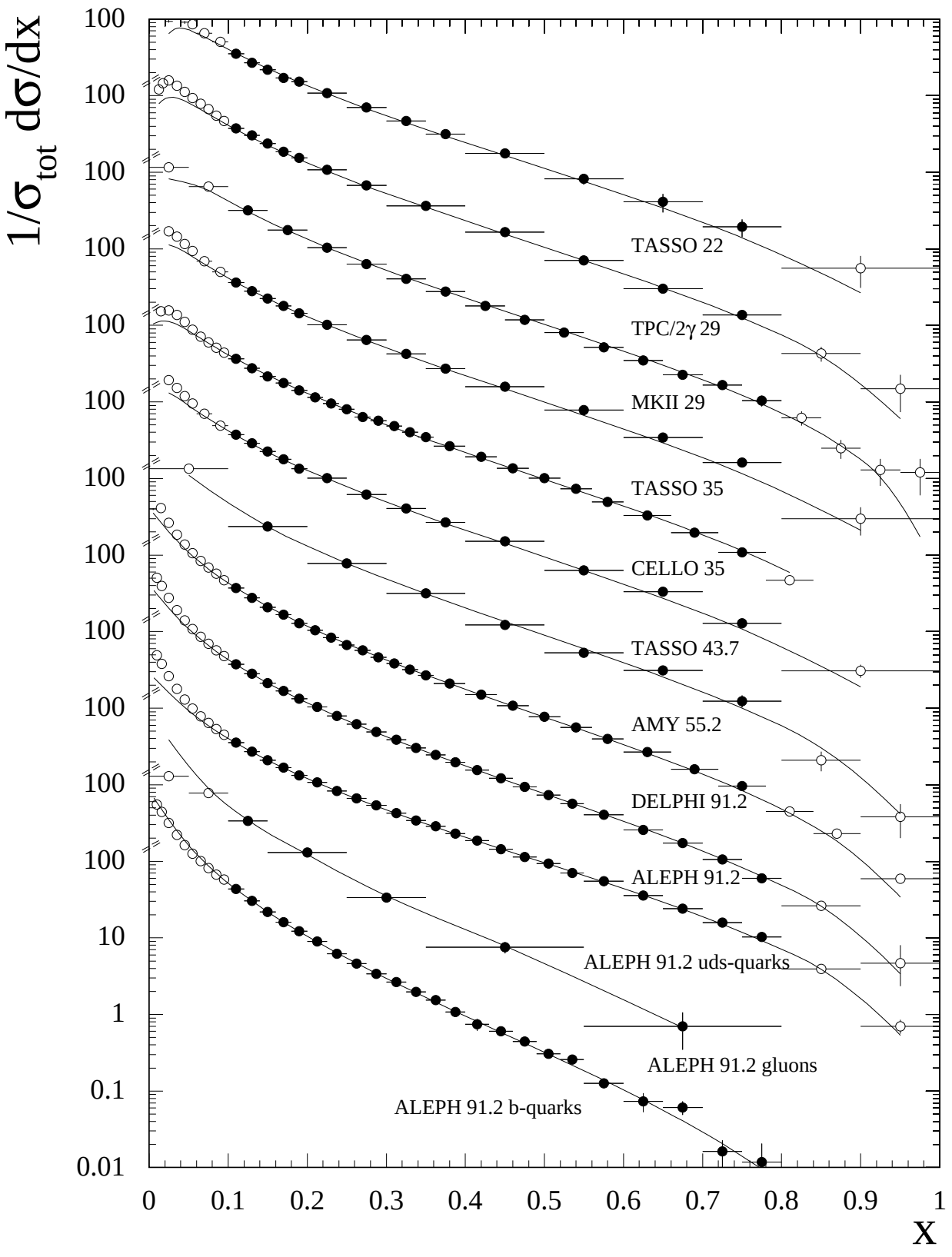

**Figure 8.** Scaling violation in $e^+e^-$ fragmentation functions.

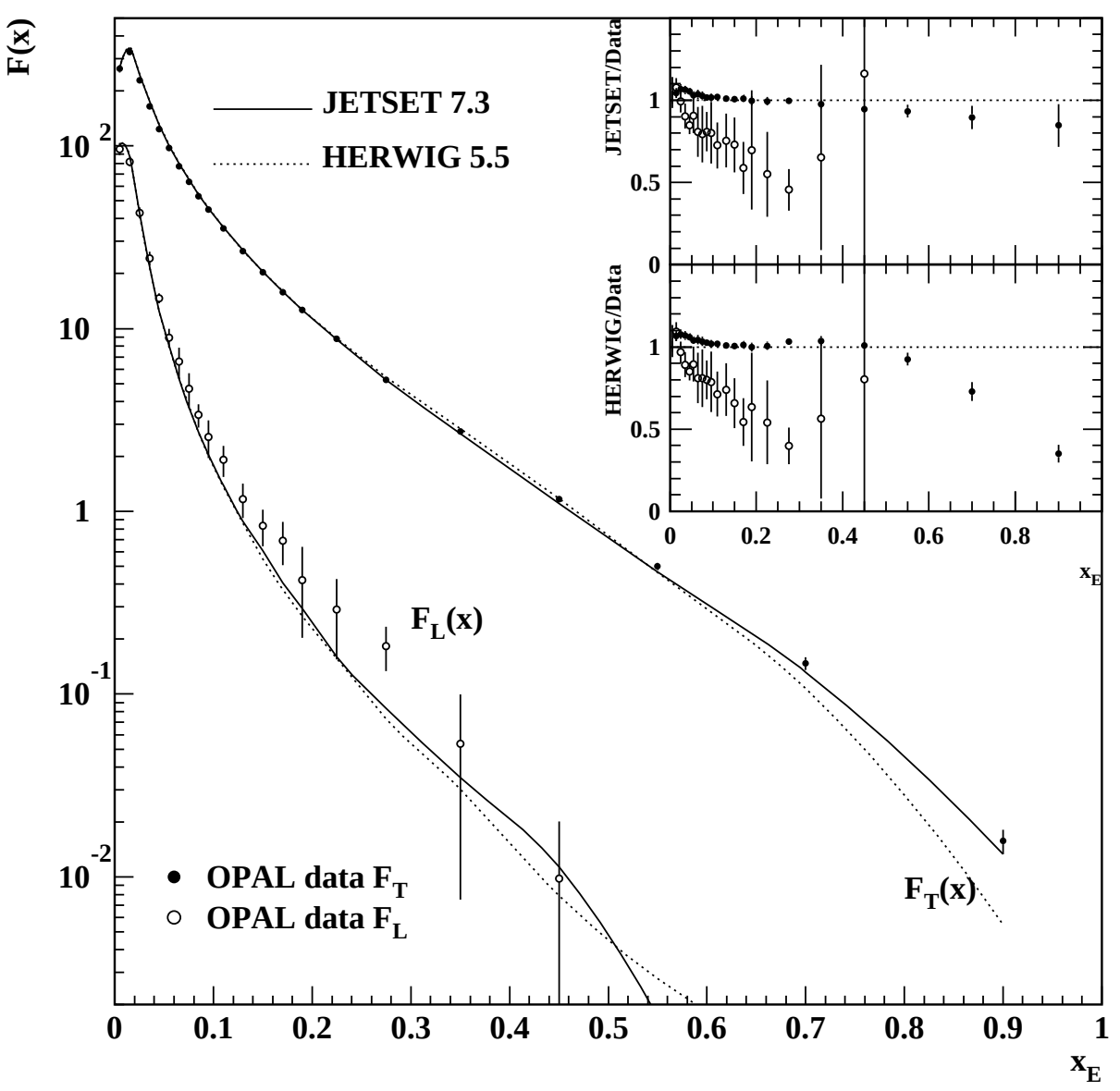

**Figure 9.** Transverse and longitudinal fragmentation functions as measured by the OPAL Collaboration.

Thus the gluon fragmentation function $D_g$ can be extracted to leading order from measurements of $F_{\rm T}$ and $F_{\rm L}$.

¶ All these functions also depend on the c.m. energy $\sqrt{s}$, which we take to be fixed here.

Figure 9 shows the OPAL data [60] on the transverse and longitudinal fragmentation functions for charged particles. Note from eq. (31) that $F_{\rm L}$ is $\mathcal{O}(\alpha_{\rm S})$ relative to $F_{\rm T}$; it also falls more steeply with $x$. Therefore even with LEP statistics the errors are large for $x > 0.3$. However, one still obtains useful information on the gluon fragmentation function over the full range of $x$, as shown in figure 10. Because the relation (31) is known only to leading order, there is an ambiguity about the energy scale at which $D_g$ is measured. Comparisons with JETSET predictions at various scales are shown.

Similar, although systematically somewhat lower, preliminary results on the longitudinal fragmentation function have been obtained by the ALEPH collaboration [45], who include this information in their fit to scaling violation as a further constraint on the gluon fragmentation function.

Summed over all particle types, the total fragmentation function satisfies the energy sum rule

$$\frac{1}{2}\int dx\, xF(x) = 1\;. \tag{32}$$

Similarly the integrals

$$\frac{1}{2}\int dx\, xF_{\rm T,L}(x) \equiv \frac{\sigma_{\rm T,L}}{\sigma_{\rm tot}} \tag{33}$$

tell us the transverse and longitudinal fractions of the total cross section. The perturbative prediction is

$$\frac{\sigma_{\rm L}}{\sigma_{\rm tot}} = \frac{\alpha_{\rm S}}{\pi} + \mathcal{O}(\alpha_{\rm S}^2)\;, \tag{34}$$

that is, the whole of the $\mathcal{O}(\alpha_{\rm S})$ correction to $\sigma_{\rm tot}$ comes from the longitudinal part. Surprisingly,

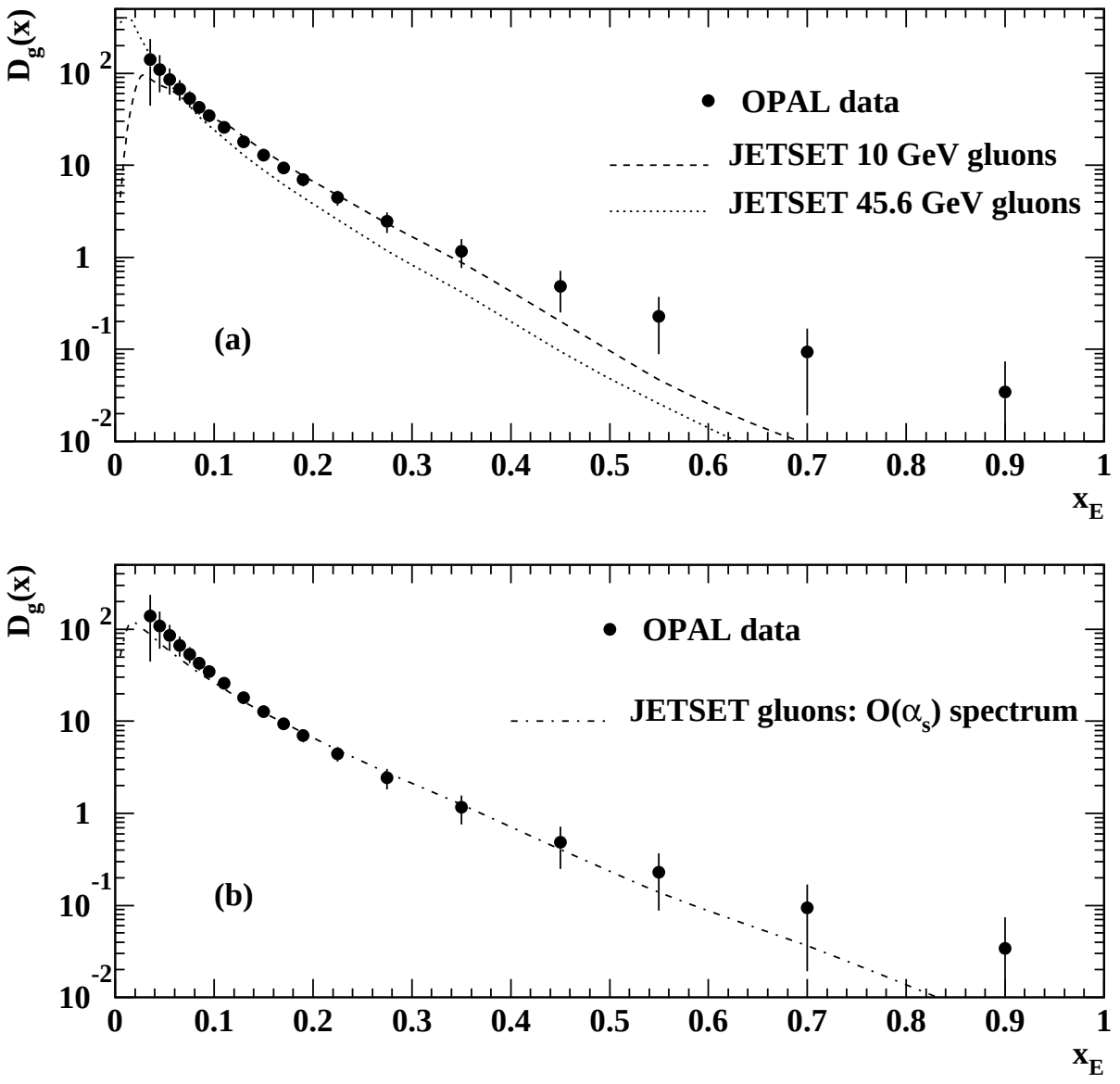

**Figure 10.** Gluon fragmentation function extracted from the OPAL data on $F_{\mathrm{T,L}}$.

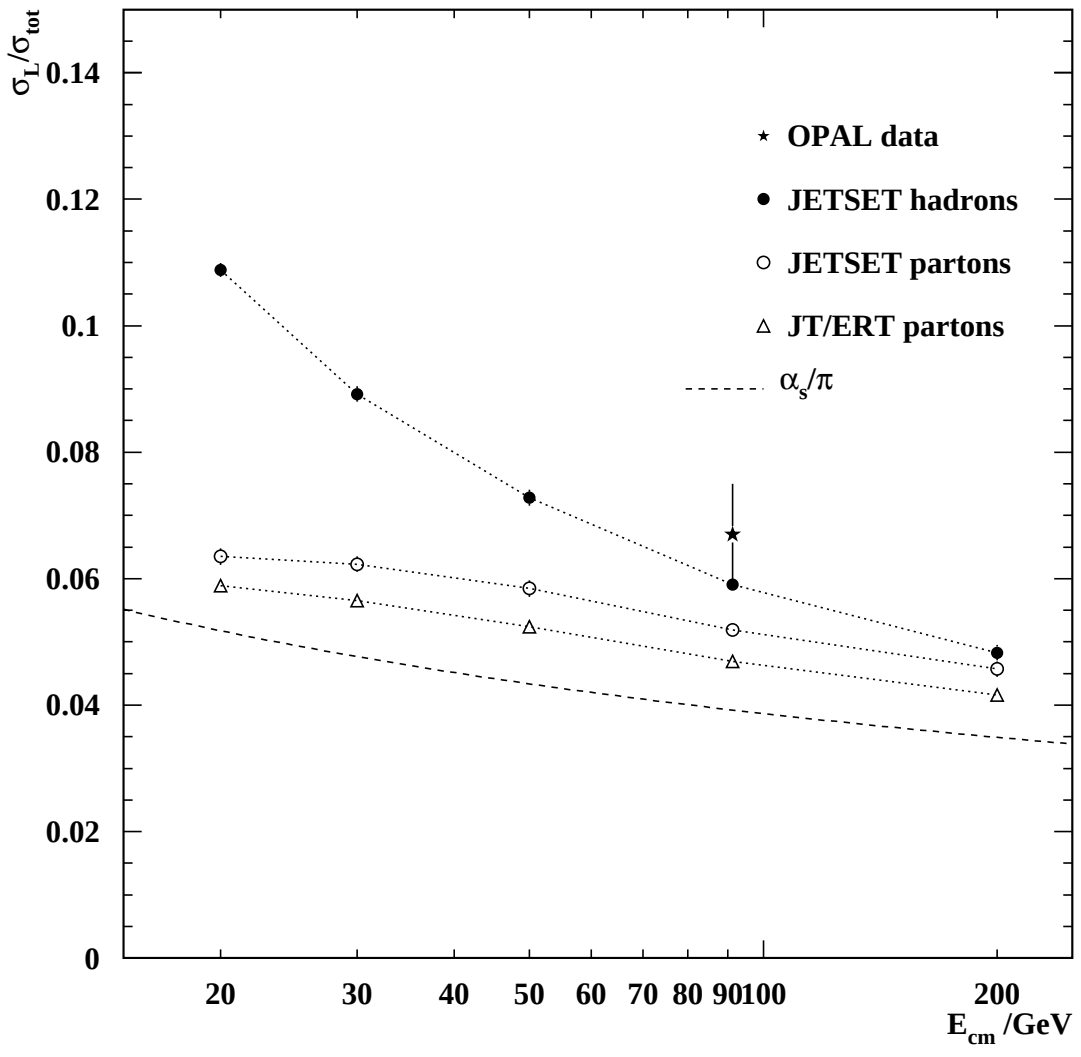

**Figure 11.** Longitudinal fraction of the total cross section as measured by the OPAL Collaboration.

the $\mathcal{O}(\alpha_S^2)$ correction has not yet been calculated. Once this has been done, eq. (34) will provide an interesting new way of measuring $\alpha_S$.

The OPAL data on $\sigma_L/\sigma_{tot}$ are shown in figure 11. It should be noted that neither of the JETSET parton-level predictions fully includes the $\mathcal{O}(\alpha_S^2)$ contribution [62]. We see, however, that the data lie well above the leading-order prediction (dashed), which suggests that higher-order and/or non-perturbative corrections are significant. An estimate of the latter is provided by the difference between the JETSET curves for hadrons and partons. This difference shows a clear $1/Q$ energy dependence, with a coefficient of about 1 GeV, illustrating the earlier remarks about power corrections to fragmentation. As we discuss next, both the magnitude and the energy dependence of the JETSET prediction are in accord with a simple model of hadronization.

### *3.3. Hadronization*

One serious problem in jet physics is the lack of any deep understanding of the hadronization process, in which the partons of perturbative QCD are converted into the hadrons that are seen in the detectors. On general grounds, we expect that hadronization and other non-perturbative effects should give rise to power-suppressed ($1/Q^p$) corrections. At present there are no solid arguments to exclude contributions with $p < 2$ for observables like event shapes and jet rates, which are not fully inclusive with respect to final-state hadrons. This is in contrast to deep inelastic structure functions and the total $e^+e^-$ hadronic cross section, where arguments based on the operator product expansion suggest that the dominant power corrections should decrease like $1/Q^2$ and $1/Q^4$ respectively [63].

From the successes of the QCD Monte Carlo programs HERWIG and JETSET, we can hope that the hadronization models built into those programs provide some guidance on the broad features of the process. What the programs suggest is that corrections to event shapes and jet rates are still substantial at energy scales $Q \sim M_Z$, typically around 10% of the leading-order QCD predictions. This is comparable with the next-to-leading $\mathcal{O}(\alpha_S^2)$ terms. Furthermore hadronization effects fall off rather slowly with increasing energy, roughly like $1/Q$. Thus it is imperative to understand hadronization better in order to reap the benefit of future $\mathcal{O}(\alpha_S^3)$ calculations of event shapes and jet rates.

One formal theoretical approach to power corrections that may be useful is the study of infrared renormalons, which are generated by the divergence of perturbation theory at high orders [63, 64]. Renormalons are associated with power-suppressed effects but it is not clear what they have to do with the hadronization process. It appears that there is a renormalon contribution to event shapes, corresponding to a $1/Q$ correction, which cancels in the total cross section [65].

The empirical magnitudes of the $1/Q$ corrections to different observables, and some features of the renormalon approach, are reproduced by the following simple "tube" model+ of the hadronization

+ This model is essentially the simplest version of the string model; we call it a tube to avoid confusion with the more sophisticated Lund string model [66].

process [67]. We suppose that the jet of hadrons from a parton of energy $E = Q/2$ occupies a tube in rapidity–transverse momentum space relative to the jet axis, with proper ("transverse") mass $\mu$ per unit rapidity. Then the energy of the jet is

$$E = \mu \int_0^Y dy \cosh y = \mu \sinh Y \qquad (35)$$

where $Y$ is the length of the tube, and so

$$Y \sim \log(Q/\mu) \ . \qquad (36)$$

On the other hand the jet momentum is

$$P = \mu \int_0^Y dy \sinh y = \mu(\cosh Y - 1) \ . \qquad (37)$$

From Eqs. (35) and (37) we expect a mean-square hadronization contribution to the jet mass of

$$\langle M^2 \rangle_{\rm had} = E^2 - P^2 \sim \mu Q \ . \qquad (38)$$

Comparing the perturbative prediction with experiment, one finds that a hadronization correction corresponding to

$$\mu \sim 0.5 \text{ GeV} \qquad (39)$$

is required. Note that this implies a fairly large jet mass, about 7 GeV at $Q \sim M_Z$, in addition to any perturbative contribution.

Given the parameter (39), hadronization corrections to other event shapes can be computed. For example the thrust of a two-jet event is

$$T = P/E \sim 1 - 2\mu/Q \ . \qquad (40)$$

Thus the hadronization correction to the thrust is expected to be

$$\langle \delta T \rangle_{\rm had} \sim -\frac{1 \text{ GeV}}{Q} \ . \qquad (41)$$

As shown in Fig. (12), this agrees well with experiment.

With a little more effort the correction to the longitudinal cross section, discussed in sect. 3.2, is found to be

$$\left\langle \frac{\delta \sigma_{\rm L}}{\sigma_{\rm tot}} \right\rangle_{\rm had} = \frac{\pi\mu}{2Q} \sim +\frac{0.8 \text{ GeV}}{Q} \qquad (42)$$

which agrees well with the JETSET prediction shown in figure 11.

## 4. Comparative jet studies

### *4.1. Comparing quark and gluon jets*

The characteristic differences between quark and gluon jets have been a matter of keen interest and hot debate since the earliest days of jet physics. Because the gluon has a larger colour charge that a quark, one expects it to radiate more gluon bremsstrahlung, leading to a jet with a higher particle multiplicity and a broader angular distribution. Asymptotically, the average multiplicities (and logarithmic angular sizes) would be expected to be in the ratio of the colour charges-squared, $g/q \sim C_A/C_F \sim 9/4$. However, QCD Monte Carlo studies suggest that this ratio is greatly reduced by higher-order and hadronization corrections at present energies, and early experimental studies saw differences of only limited significance [68].

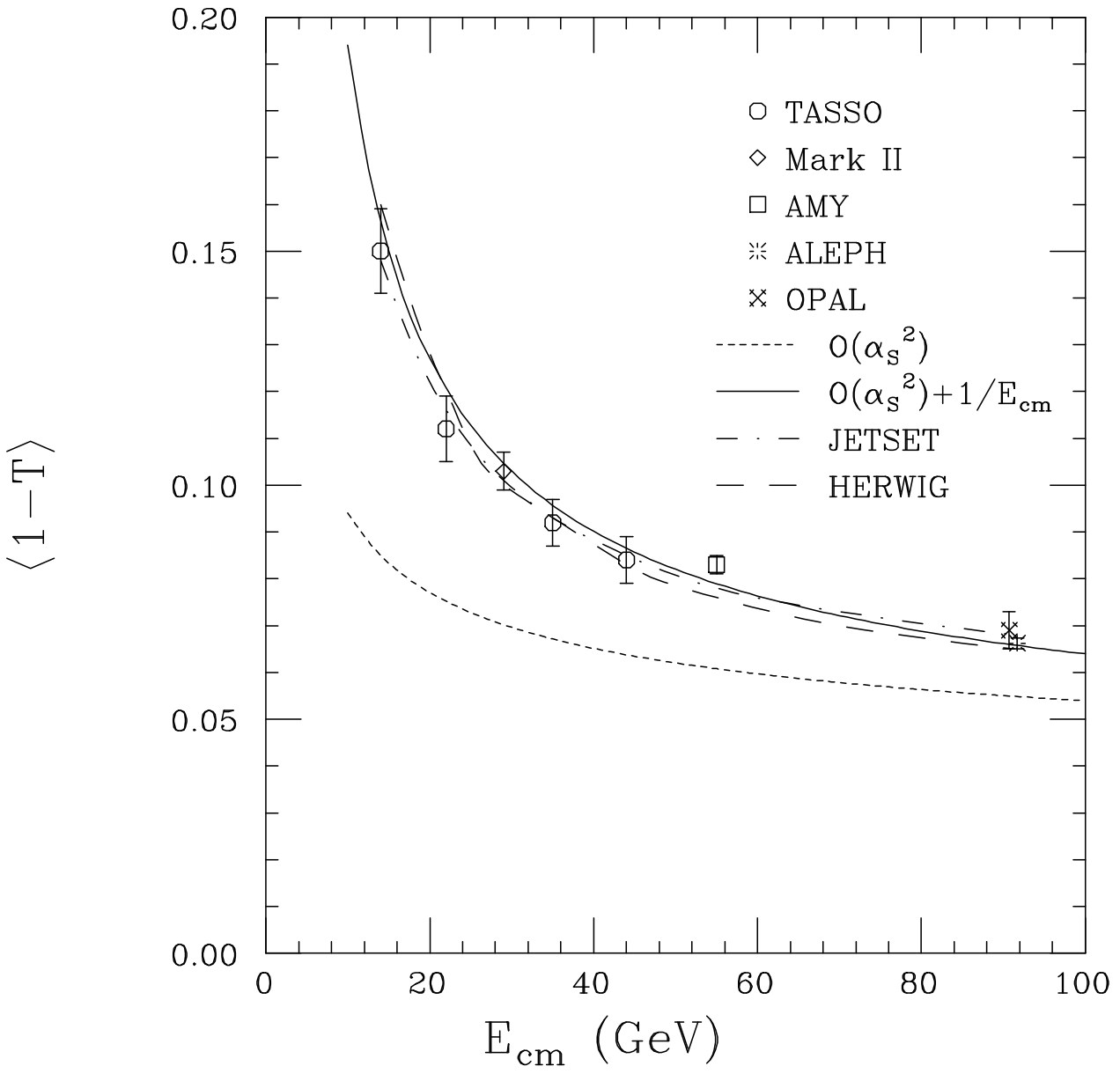

**Figure 12.** Mean value of 1 − thrust in $e^+e^-$ annihilation, as a function of centre-of-mass energy.

The experimental study of quark-gluon differences has undergone great progress in recent years, thanks to the development of silicon vertex detectors. With these devices one can tag heavy quark jets, by identifying secondary vertices due to heavy flavour decays. The earliest approach, pioneered by the OPAL collaboration [69], was to select a two-fold symmetric subset of three-jet events in which one of the two lower-energy jets is vertex tagged. The highest-energy jet is most unlikely to be a gluon and so the untagged one of the lower-energy jets is gluon-enriched. This demonstrated clear differences in multiplicity and the longitudinal and transverse structure of quark and gluon jets.

Making use of the increased statistics, one may now select a more fully symmetric three-jet sample in which two jets are tagged, so that the third jet is almost certainly a gluon. Gluon jet samples with purity $\sim$ 95% can be prepared in this way. Comparing the gluon jets with the whole three-jet sample, which consists of one-third gluons and two-thirds quarks with the usual flavour mix, one can construct distributions for quark jets by subtraction. This 'double vertex' tagging method yields samples of quark and gluon jets with very similar energies and angular separations.

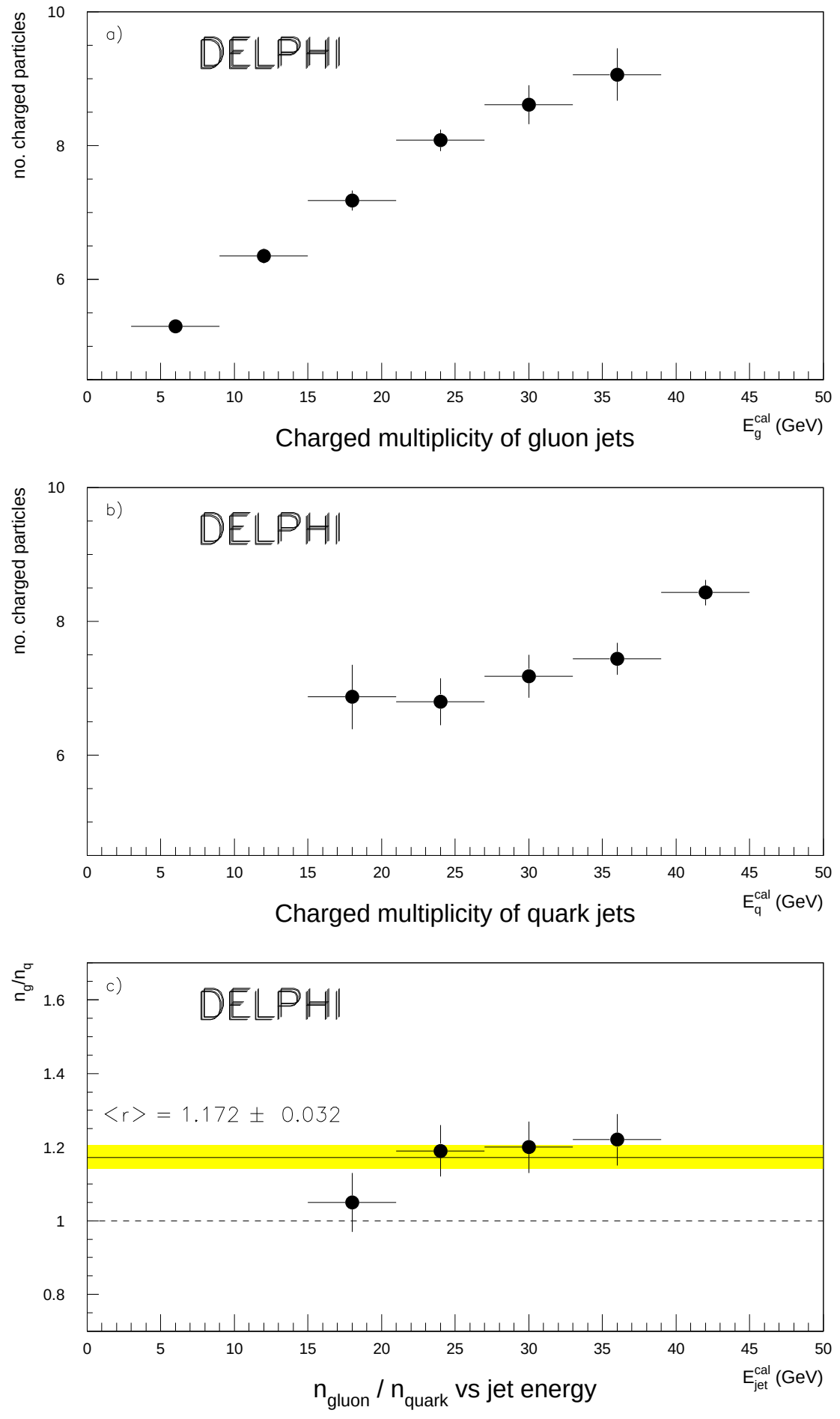

**Figure 13.** Charged multiplicities in quark and gluon jets.

Figure 13 shows preliminary results obtained by the DELPHI collaboration [70] on the charged multiplicity of quark and gluon jets using the double vertex tagging method. Figure 14 shows corresponding results on the leading particle rapidity and normalized transverse momentum-squared. In each case clear differences are seen in the direction expected. These results confirm those obtained by OPAL using single vertex tagging [69]. Similar results have been presented by ALEPH [71].

The differences between quark and gluon jets, although now shown clearly by experiment, remain small in comparison with the asymptotic expectations outlined above. For example, the charged multiplicity ratio is found to be in the region 1.2 – 1.3, depending on jet definition and energy, instead of 9/4.

The reasons for this apparent discrepancy can be illuminated by studies of the sub-jet multiplicity inside jets [72]. This is defined by first selecting a jet sample using the Durham/$k_\perp$ clustering algorithm with resolution $y_{\rm cut} = y_1$ and then counting the number of sub-jets at $y_{\rm cut} = y_0 < y_1$. Recall that the Durham/$k_\perp$ criterion for resolving two objects $i$ and $j$ at resolution $y_{\rm cut}$ is $y_{ij} > y_{\rm cut}$ where

$$y_{ij} = 2\min\{E_i^2, E_j^2\}(1-\cos\theta_{ij})/s \sim k_{\perp ij}^2/s\ . \quad (43)$$

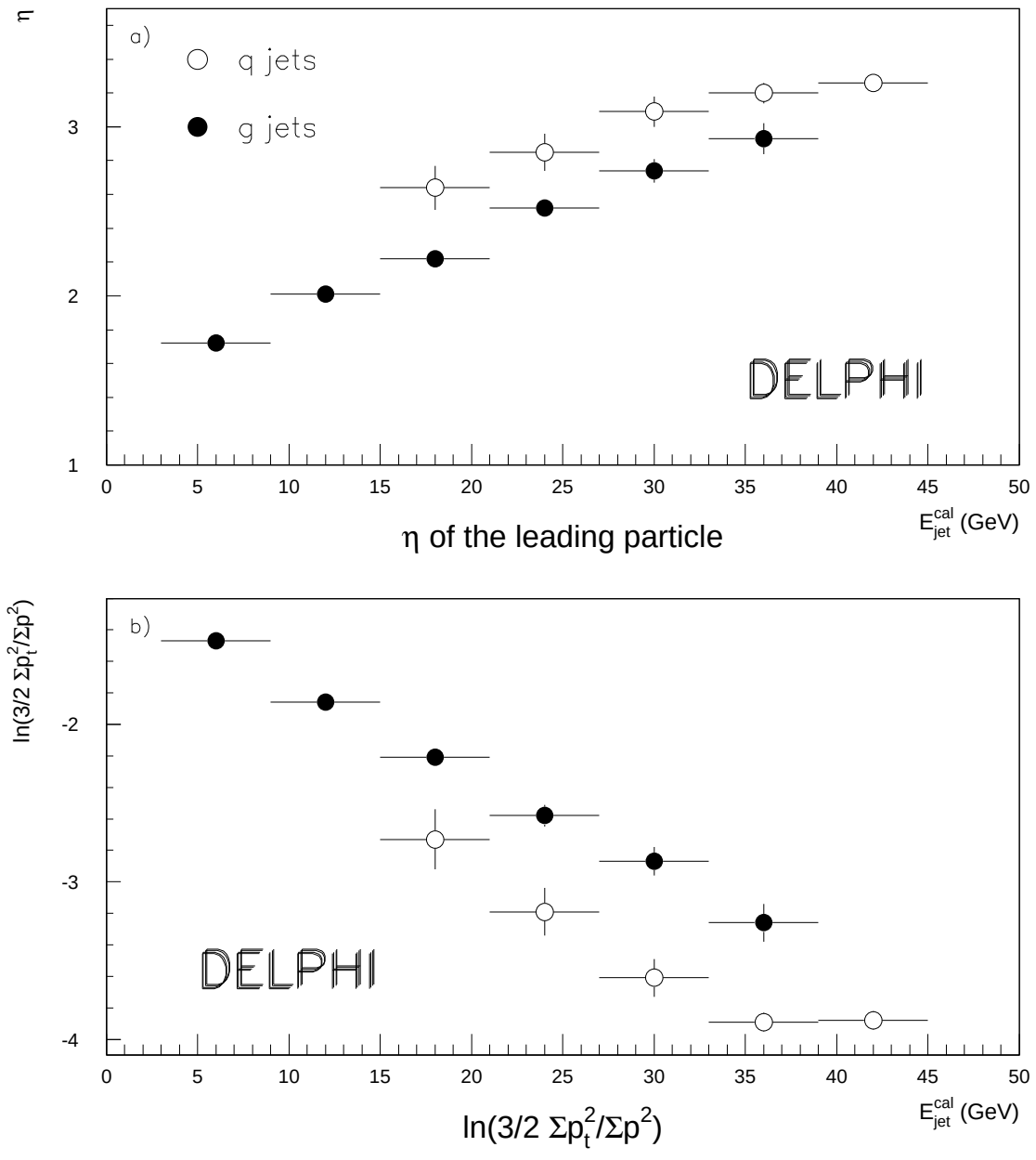

**Figure 14.** Quark-gluon jet differences as measured by the DELPHI Collaboration.

By varying $y_0$ downwards from $y_1$ we effectively sweep through the development of a jet, from the formation scale to the scale at which individual hadrons are resolved as sub-jets ($y_{\rm cut} \sim 10^{-5}$ at $\sqrt{s} = M_Z$).

Figure 15 illustrates results obtained by OPAL [73] on the average sub-jet multiplicity $M_2$ in two-jet and $M_3$ in three-jet events, without tagging. Here one expects asymptotically (for $y_0 \ll y_1 \ll 1$) that

$$\frac{M_3-3}{M_2-2} \sim 1 + \frac{C_A}{2C_F} = \frac{17}{8}\ , \quad (44)$$

because the additional jet in the three-jet events is due to a gluon. For increased sensitivity the number of extra sub-jets $M_k - k$ is counted in each case, since trivially $M_k \to k$ as $y_0 \to y_1$.

One sees from figure 15 that two phenomena combine to keep the ratio (44) well below its asymptotic value. At high $y_0$ (early in the evolution of the jets) there is a strong QCD coherence effect that suppresses the ratio near $y_0 = y_1$. This can be computed to all orders in next-to-leading logarithmic approximation (NLLA) when $y_0 \lesssim y_1 \ll 1$, yielding the solid curves. The first-order NLLA result, which

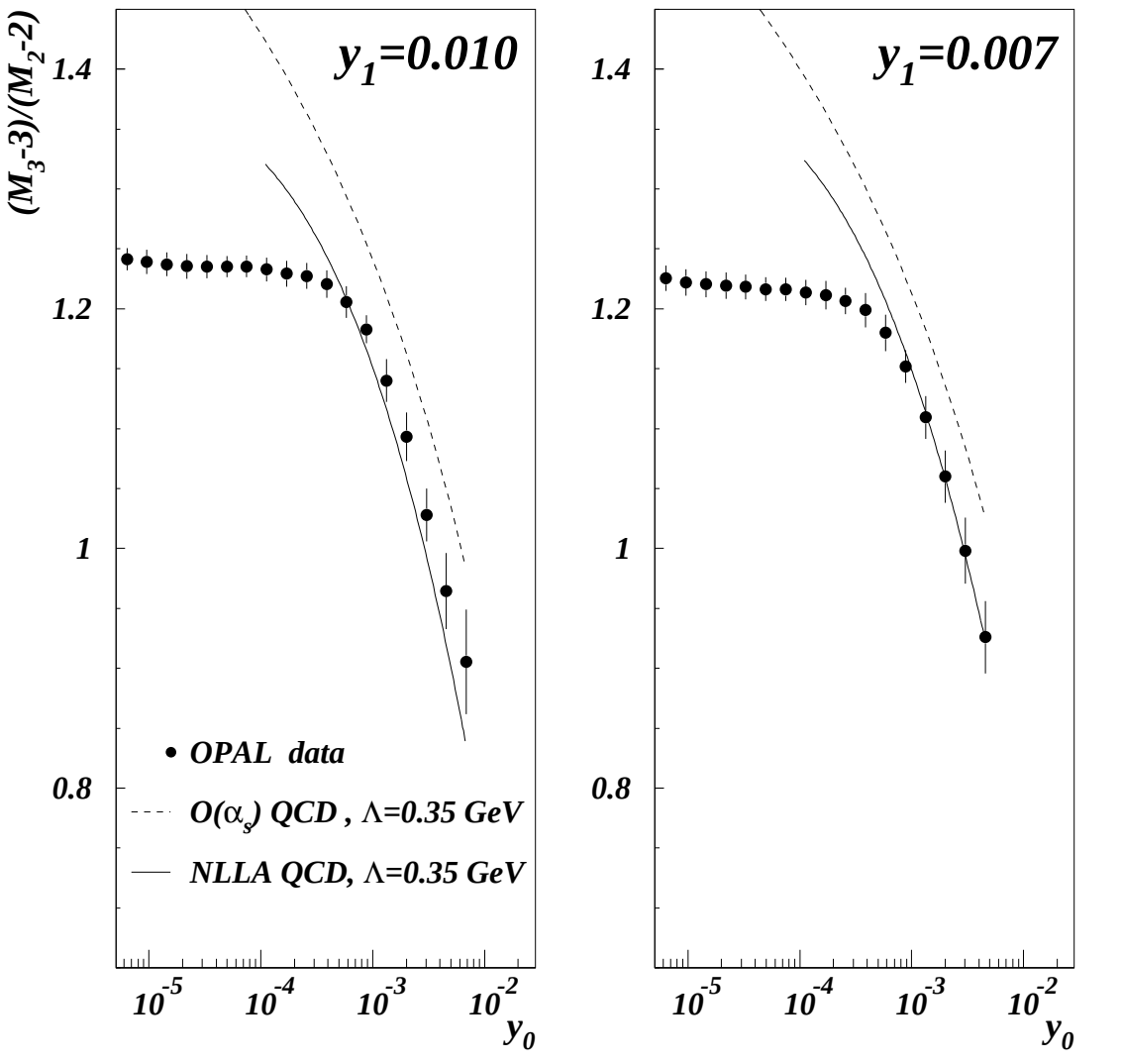

**Figure 15.** Ratio of sub-jet multiplicities in two- and three-jet events as measured by the OPAL Collaboration.

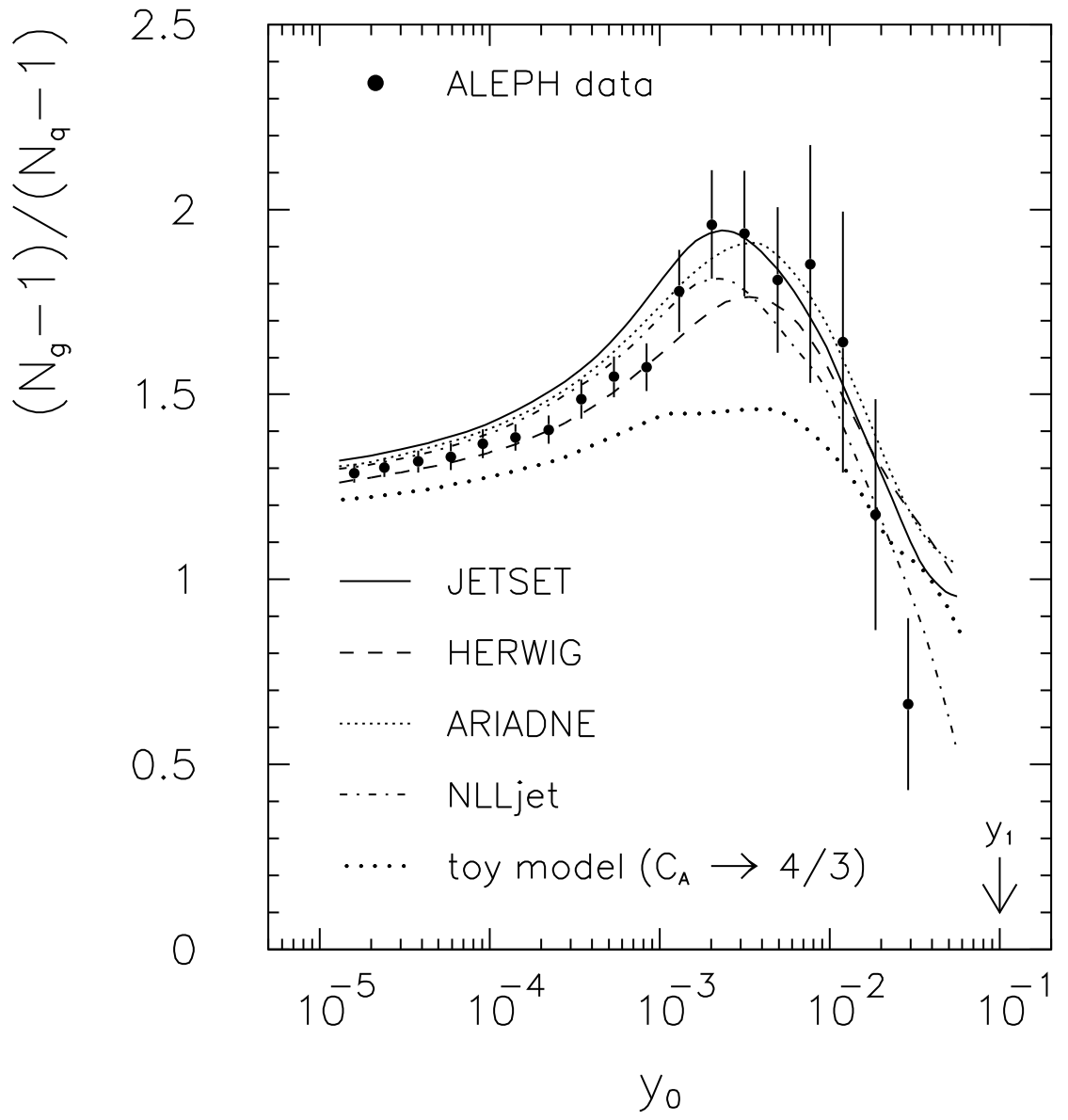

**Figure 16.** Ratio of subjet multiplicities in quark and gluon jets as measured by the ALEPH Collaboration.

is not far from the full $\mathcal{O}(\alpha_S)$ prediction shown by the dashed curves, is [72]

$$\frac{M_3-3}{M_2-2} \sim 1+\frac{C_A(L_0-\frac{1}{3}L_1-\frac{14}{3})+\frac{2}{3}N_f}{2C_F(L_0+L_1-3)} \qquad (45)$$

where $L_{0,1}=\ln(1/y_{0,1})$. The suppression arises from a restriction on additional gluon radiation due to angular ordering. As $y_0$ decreases, the restriction becomes less important and the multiplicity ratio rises rapidly. But then it suddenly flattens off, at a value below 1.3, when $y_0 \sim 4\times 10^{-4}$.

The flattening appears to be a hadronization effect. The corresponding relative transverse momentum of two just-resolved sub-jets is $\sqrt{y_0 s} \sim$ 2 GeV. This may seem large, but we have to remember that hadronization generates about 500 MeV of transverse energy per unit rapidity, and so a jet occupying four units of rapidity will be resolved into two sub-jets at this scale.

To see a larger increase in the sub-jet multiplicity we need a bigger interval of $y_0$ between the jet scale $y_1$ and the hadronization scale $y_{\rm had} \sim 4$ GeV$^2/s$. At fixed $s$ this can only be arranged by increasing $y_1$. We do indeed see a small increase in figure 15 when $y_1$ is increased from 0.007 to 0.01. The ALEPH collaboration [74] have carried this idea further, choosing $y_1 = 0.1$. Here the three-jet events are extremely symmetrical and by double-tagging one can compare sub-jet multiplicities in quark and gluon jets of the same energies, with the results shown in figure 16. We see that in the relatively large interval of jet evolution now available the multiplicity ratio comes close to the asymptotic value of 9/4 when $y_0 \sim 2\times 10^{-3}$, but it is then dragged down, presumably by hadronization, to the value around 1.3 observed at the hadron level.

There is still much to be learnt about why the differences between quark and gluon jets are not larger. The study of sub-jets provides a useful new tool with which to probe this phenomenon. By varying the sub-jet resolution $y_0$ one is able to follow the development of a jet and to observe the onset of hadronization. It would be interesting to apply this type of analysis to enriched quark and gluon samples obtained in other processes, such as hadron-hadron collisions.

### 4.2. *Comparing $e^+e^-$ and $p\bar{p}$ jets*

The comparative study of jets produced in different reactions in another area in which quantitative progress is being made. Over the years it has become clear that the properties of jets depend on the precise operational procedure by which they are defined and found. For meaningful comparisons one therefore needs a procedure that can be applied identically to different processes. In $e^+e^-$ annihilation, and to some extent in lepton-hadron scattering, the preferred approach has been the use of jet clustering procedures, whereas in hadron-hadron collisions the use of cone jet algorithms has been predominant.

There is no reason why jet clustering should not be extended to hadron-hadron collisions [75], but for the moment that has not been done, and so instead the OPAL collaboration [43] have applied a cone algorithm, essentially identical to that used by the Tevatron $p\bar{p}$ collider experiments [76, 77], to their $e^+e^-$ data.

In a hadron collider, the most useful kinematic variables are the transverse energy $E_T = E\sin\theta$ and the pseudorapidity $\eta = -\ln\tan\theta/2$, where $\theta$ is the angle with respect to the beam direction. Then a cone jet is defined by a certain minimum amount of $E_T$ inside a region

$$(\Delta\eta)^2 + (\Delta\phi)^2 < R^2 \qquad (46)$$

where $\phi$ is the azimuthal angle around the beam direction. Such a definition is natural for hadronic collisions, which involve incoming partons with a wide range of longitudinal momenta, but not for $e^+e^-$ annihilation, where the hard process c.m. frame is known. Nevertheless the same definition can be adopted, for precise comparability. *

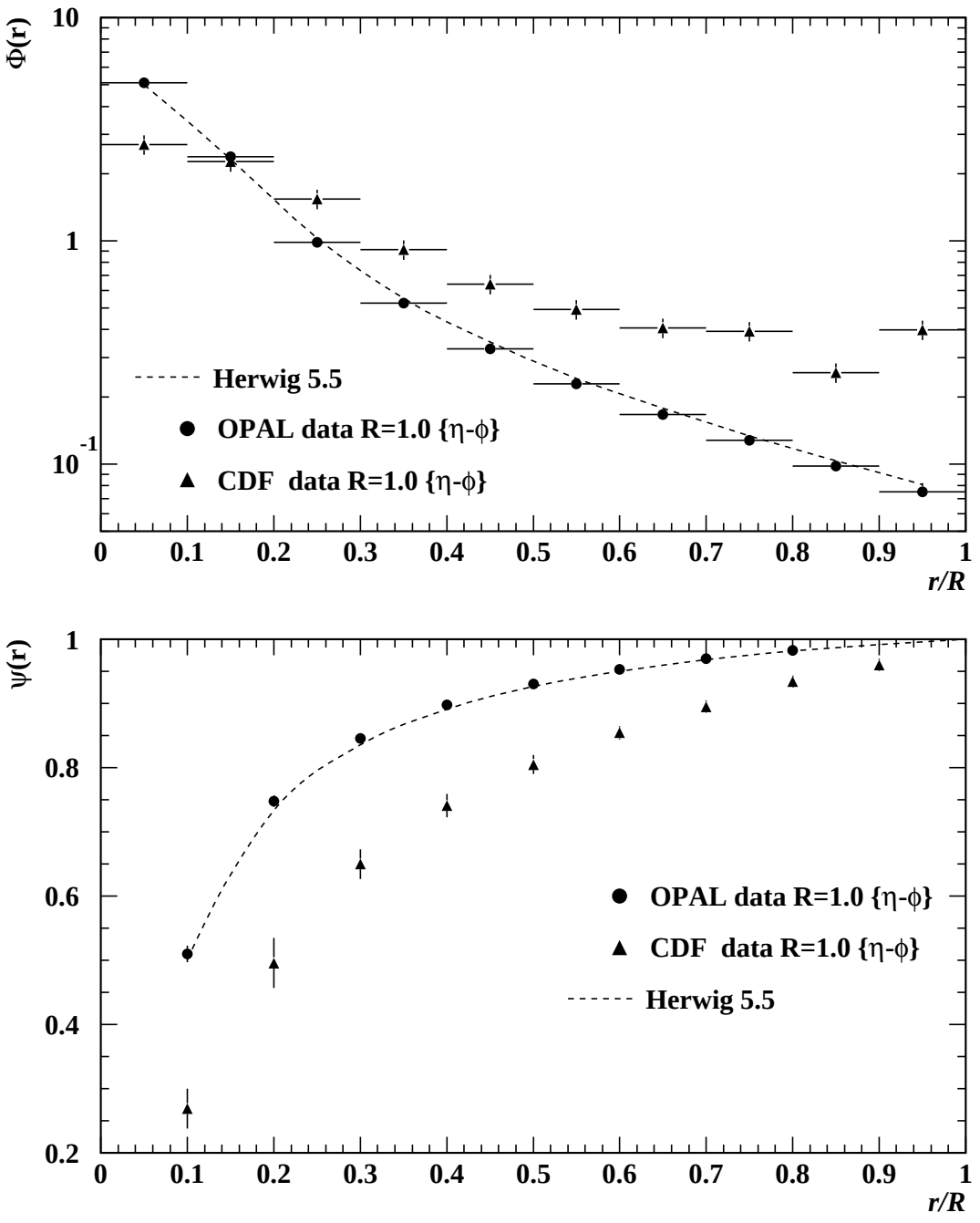

**Figure 17.** Energy flow profiles of jets, measured with identical cone algorithms in $e^+e^-$ and $p\bar{p}$ collisions.

The results on the energy profiles of jets with $E_T(R=1) > 35$ GeV are shown in figure 17. Here $\Psi(r)$ is the fraction of $E_T$ lying in a sub-cone of size $r < R$ and $\Phi(r) = R d\Psi/dr$ is the corresponding differential distribution. We see that the OPAL $e^+e^-$ jet profile falls more steeply than the $p\bar{p}$ one obtained by CDF: the $e^+e^-$ jets are narrower. Monte Carlo studies suggest that the main reason for this is that the $p\bar{p}$ jets at these relatively low $E_T$ values are mostly gluon jets, which, as discussed above, should indeed be broader than the $e^+e^-$ quark jets.

Experimental support for this interpretation of figure 17 is provided by preliminary data from D0 [77] on the comparative profiles of forward and central jets (figure 18). In the forward directions, hard collisions involving fast constituents of the incoming hadrons are selected, and these tend to be quarks more often than the slower constituents involved in central collisions. Thus the narrower profile seen in forward jets is again associated with the predominance of quarks over gluons.

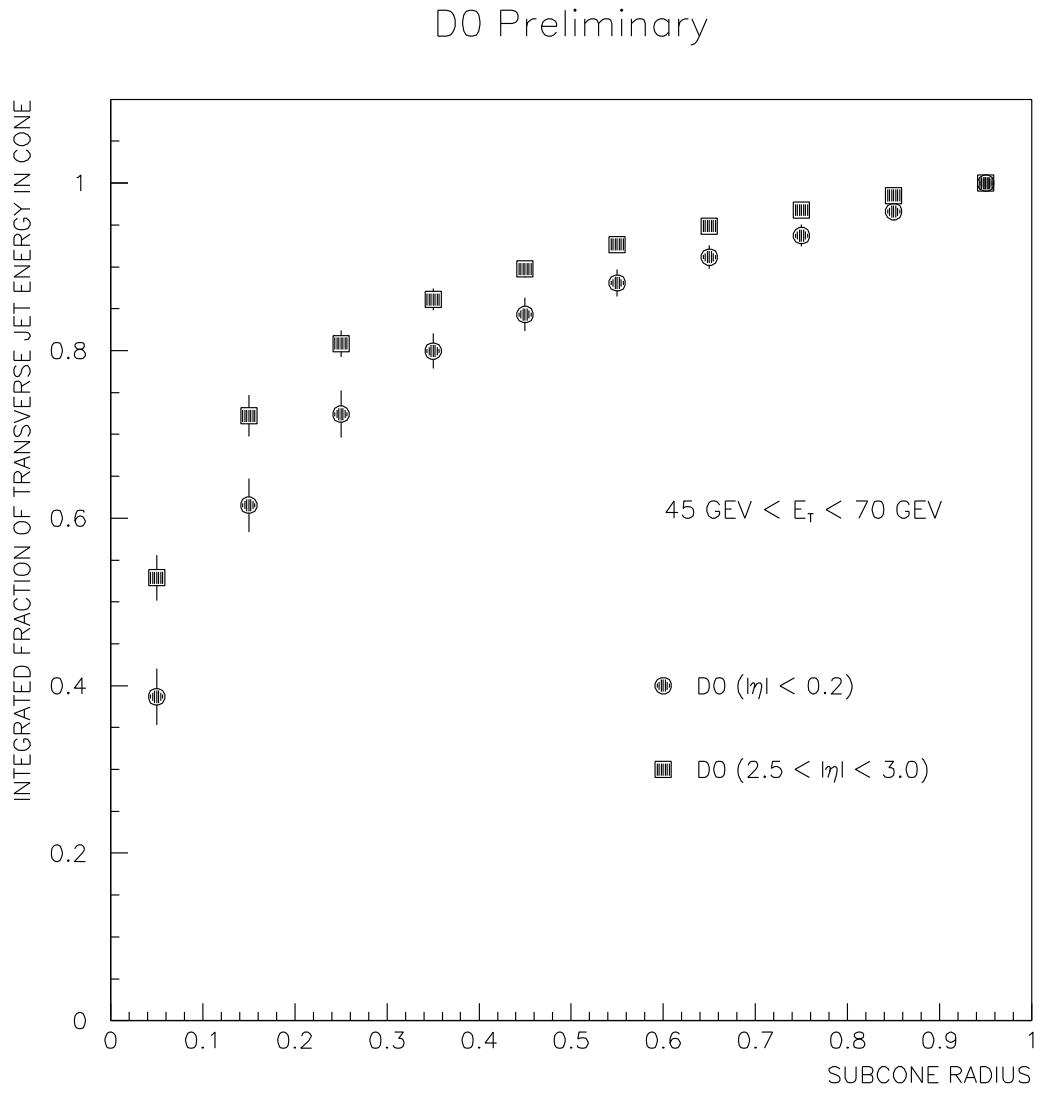

**Figure 18.** Energy flow profiles of central and forward jets, as measured by the D0 Collaboration.

### 4.3. *Comparing $e^+e^-$ and ep jets*

Jet physics at HERA promises to be a rich field of study. We have already discussed the jet rate measurement as a method for $\alpha_S$ determination. Here I would like to mention a way of looking at jet fragmentation which permits detailed comparisons between $e^+e^-$ and deep inelastic ep collisions without the necessity of using a jet-finding algorithm.

The idea is simply to look at the "current jet hemisphere" of deep inelastic scattering *in the Breit frame of reference* [78]. The Breit frame is the one in which the momentum transfer $q^\mu$ has only a $z$-component: $q^\mu = (0,0,0,-Q)$. In the parton model, in this frame the struck quark enters from the left with $z$-momentum $p_z = \frac{1}{2}Q$ and sees the exchanged virtual boson as a "brick wall", from which it simply rebounds with $p_z = -\frac{1}{2}Q$. Thus the left-hand hemisphere of the final state should look just like one hemisphere of an $e^+e^-$ annihilation event at $E_{\rm cm} = Q$. The right-hand ("beam jet") hemisphere, on the other hand, is different from $e^+e^-$ because it contains the proton remnant, moving with momentum $p_z = (1-x)Q/2x$ in this frame.

Figure 19 shows the preliminary results of such a comparison by the ZEUS collaboration, in this case comparing the charged multiplicity in the Breit frame current hemisphere with that in $e^+e^-$ annihilation. We see that the two sets of data do seem to follow the same curve.

* In fact the OPAL collaboration find that their conclusions are unaffected if they define the $e^+e^-$ cones more naturally, in terms of energy and solid angle.

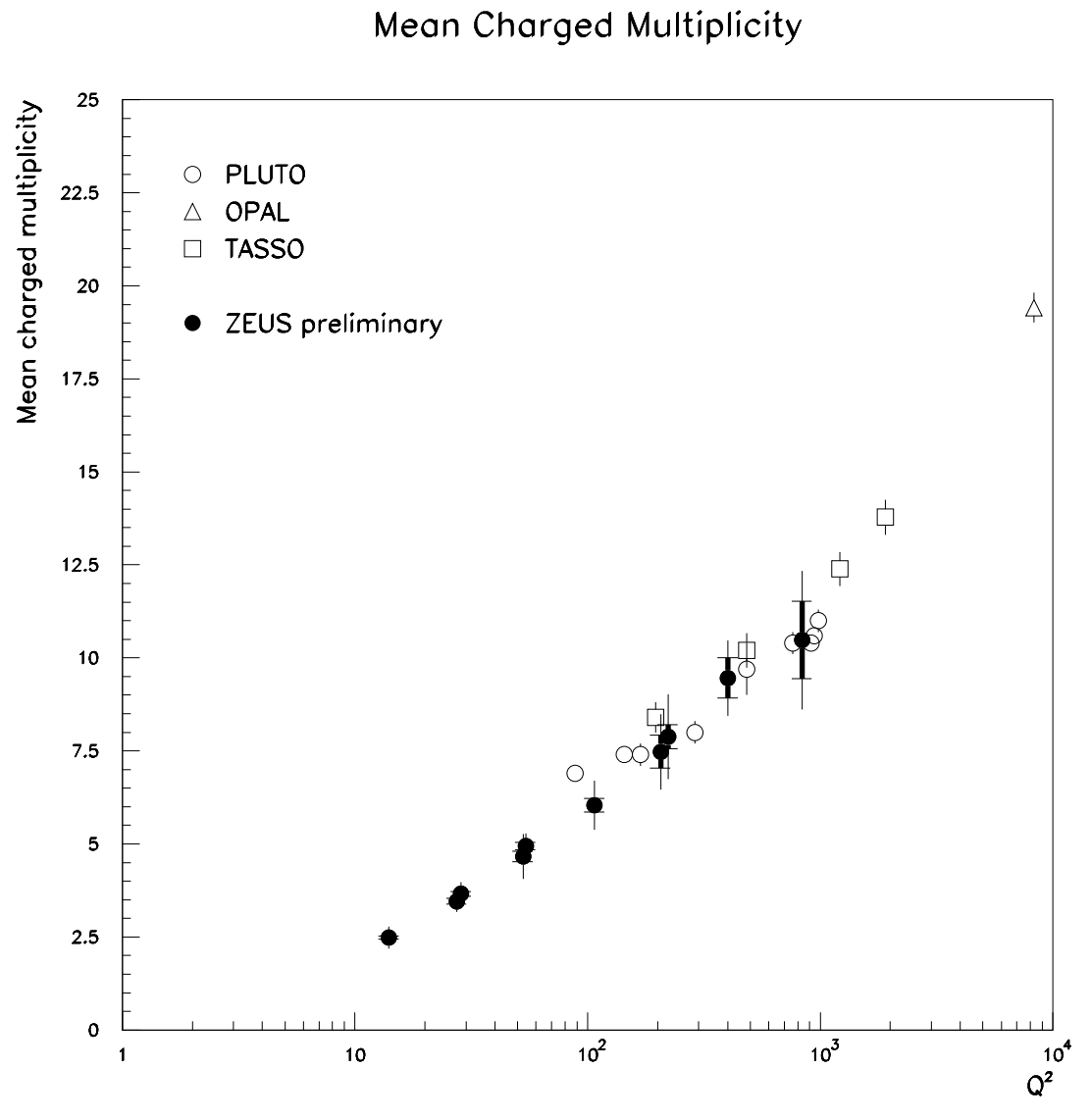

**Figure 19.** Mean charged multiplicity in ep current jet fragmentation as a function of $Q^2$, compared with that in $e^+e^-$ annihilation at $E_{\rm cm} = Q$.

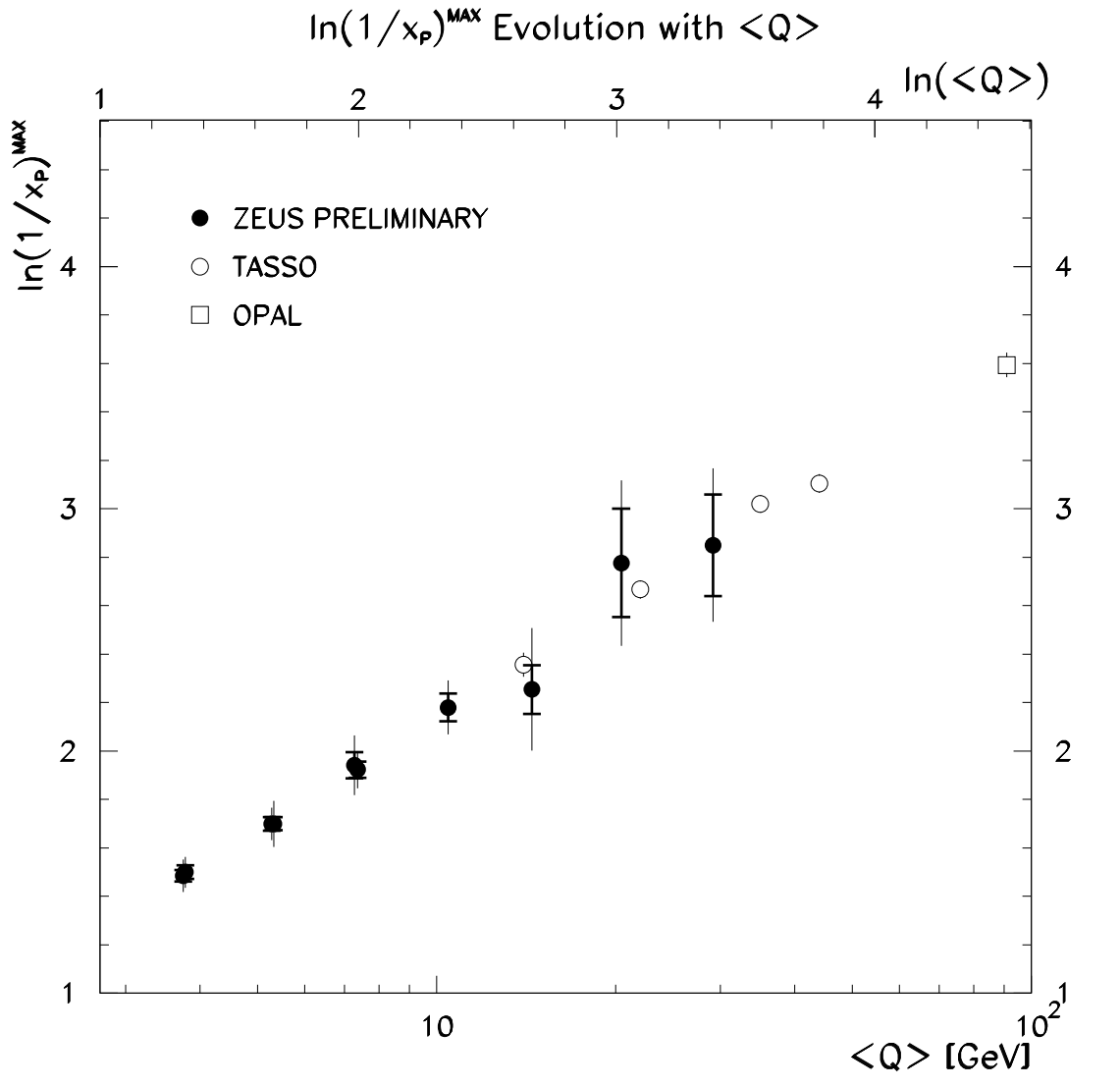

**Figure 20.** Position of the peak in $\ln(1/x_p)$ in ep current jet fragmentation, compared with that in $e^+e^-$ annihilation at $E_{\rm cm} = Q$.

A more elaborate comparison is shown in figure 20. Here the position of the peak in the distribution of $\ln(1/x_p)$, where $x_p = 2p/Q$ in the Breit frame, is compared with that found in $e^+e^-$ annihilation. We see again that the preliminary ep data follow the $e^+e^-$ curve. The QCD prediction is that the peak position should vary roughly as $\frac{1}{2}\ln Q$, which is consistent with the data.

## 5. Other topics

### 5.1. *Colour factor analyses*

Perhaps even more fundamental than the coupling constant of QCD is its underlying colour structure. The fact that the colour group is SU(3), and that the quarks belong to the fundamental representation, fixes the relative magnitudes of all the couplings of the theory. The first non-trivial test of these relative magnitudes in $e^+e^-$ annihilation comes from the distributions of four jet final states. Roughly speaking, the contribution of the gluon splitting process $g \to gg$, relative to gluon bremsstrahlung, is determined by the ratio of quadratic Casimir operator values $C_A/C_F = 9/4$ in SU(3). Similarly the relative contribution of the $g \to q\bar{q}$ and bremsstrahlung processes is fixed by $T_R/C_F = 3N_f/8 = 15/8$ for 5 fully active flavours.

In four-jet studies [79, 80], the contributions of different processes are not distinguished, but rather the angular correlations between the jets are compared with perturbative predictions to extract the colour factor ratios. A disadvantage of this type of analysis is that only tree-level predictions are available, and there is no way of assessing the effects of higher orders.

A new approach [81] is to use the fact that the two- and three-jet cross sections are also sensitive to the colour factor ratios via higher-order corrections. Since the dependence appears first at $\mathcal{O}(\alpha_S^2)$, this is again effectively only a tree-level measurement, but it serves as a valuable cross-check on the four-jet results.

Another new idea [82] is to use event shape distributions to measure the colour factors. Here again the dependence appears first at $\mathcal{O}(\alpha_S^2)$, but now there are resummed predictions available for certain event shapes in the two-jet region. The all-orders resummation gives increased sensitivity to the colour factors (although still not as great as that of the four-jet studies), and provides some indication of the effects of higher orders via renormalization scale dependence.

A summary of results, including also those from earlier hadron collider analyses, is shown in figure 21 [56]. We see there is very good overall agreement with the QCD values.

Recalling the discussion of a possible light gluino in sect. 2.4, one should note that production of such an object would change the effective value of $T_R/C_F$ to 3, which is disfavoured by the data (LEP average $= 1.52 \pm 0.74$).

### 5.2. *Prompt photon production*

The prompt production of photons at large transverse momenta in hadron collisions is a hard scat-tering process with a number of interesting features. The identification of photons is cleaner than that of other final-state particles, and, photons being truly elementary particles, their production is in a sense more direct and fundamental. The dominant subprocess in leading order is Compton-like scattering, $gq \to \gamma q$, and so one obtains useful information on the gluon distribution in the incident hadrons. Experimental data from high-energy $p\bar{p}$ colliders [83, 84] probe the region of momentum fractions in which gluons predominate.

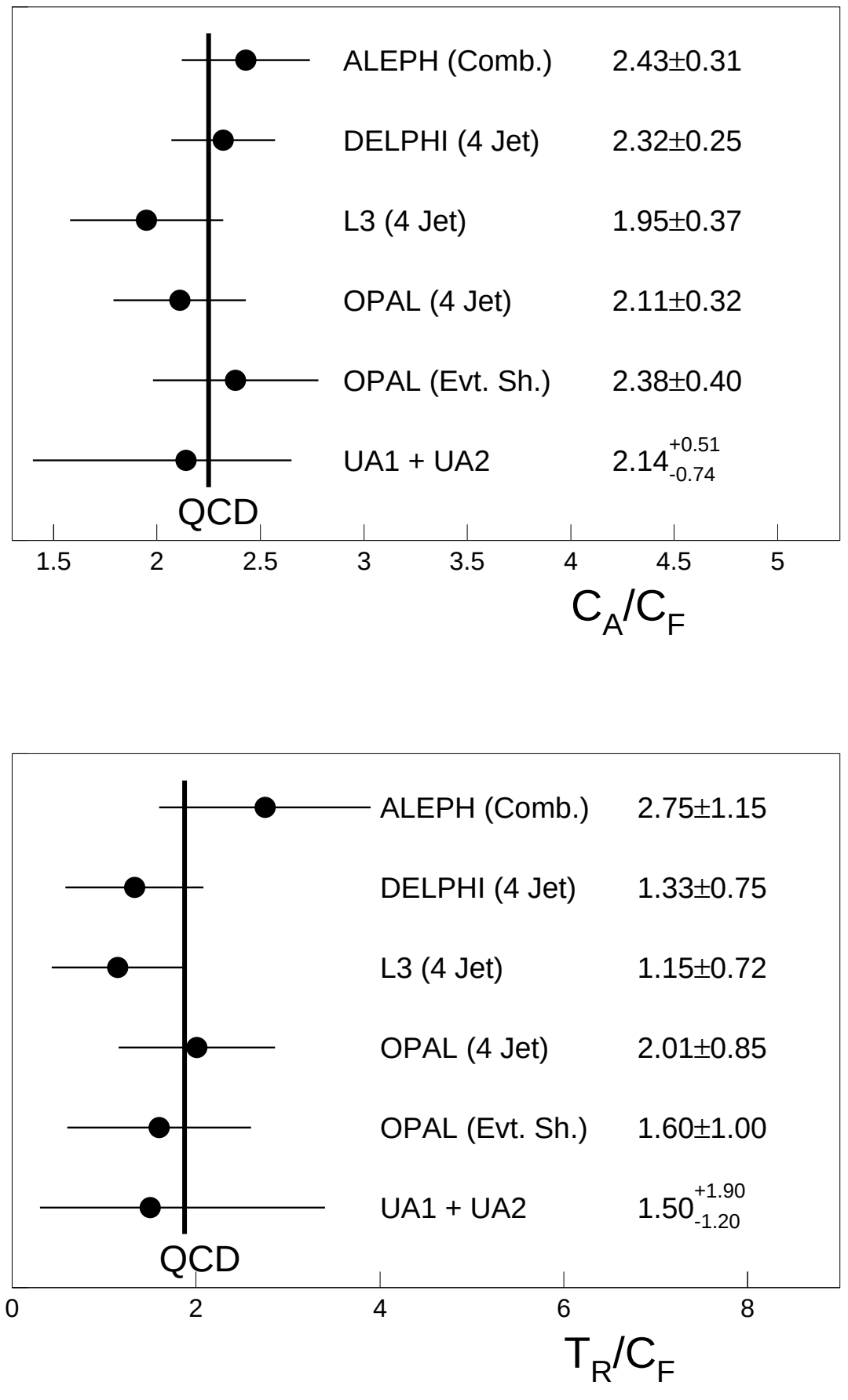

**Figure 21.** Summary of colour factor measurements.

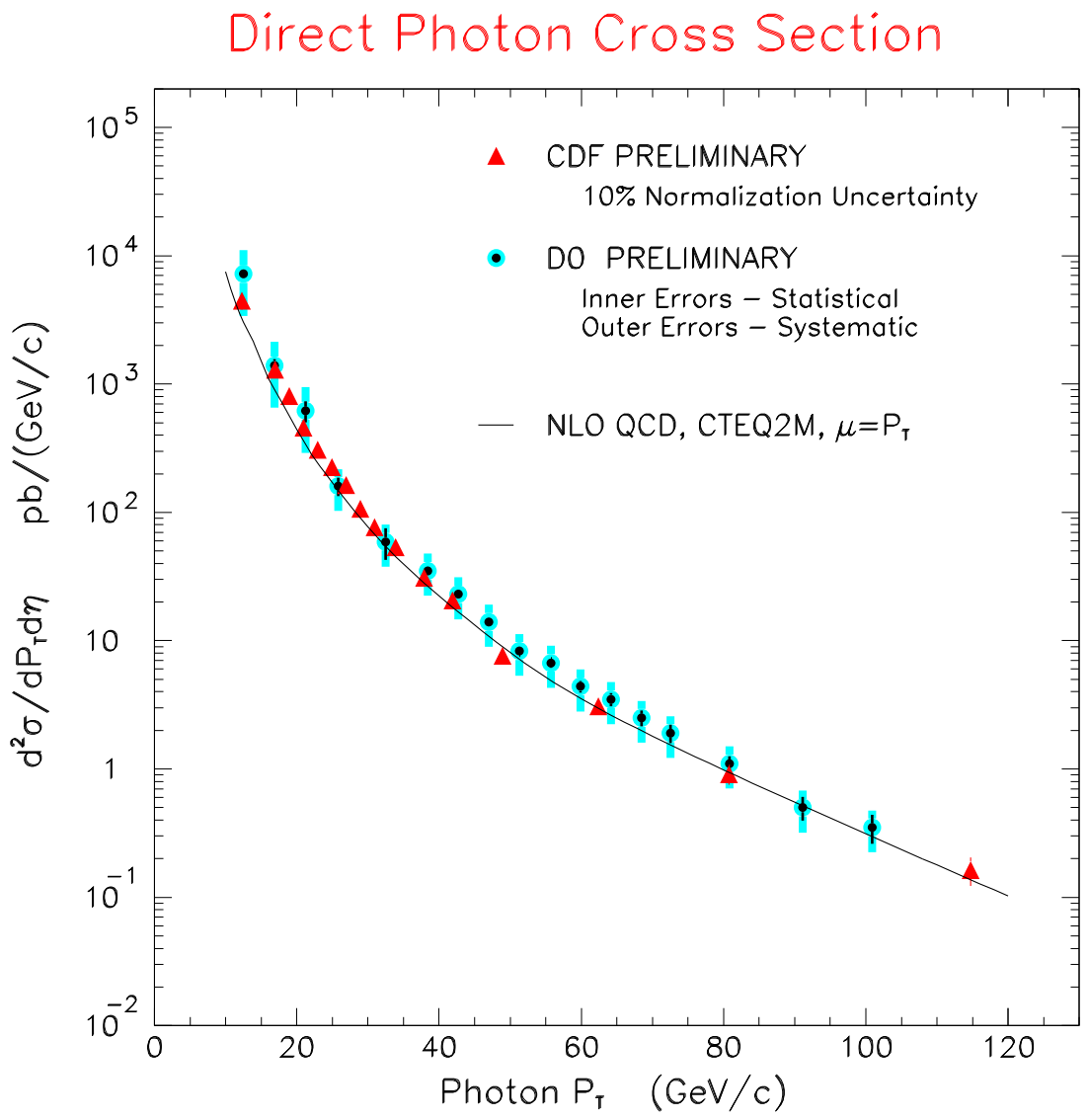

**Figure 22.** Cross section for prompt photon production at $\eta = 0$ in $p\bar{p}$ collisions at $\sqrt{s} = 1.8$ GeV.

In QCD calculations of prompt photon production one has to take account of direct hard scattering processes [85] and also fragmentation or bremsstrahlung processes [86], in which a produced parton emits a photon. In the latter case one needs parametrizations of the non-perturbative functions describing the fragmentation of partons into photons [87, 88], as mentioned in sect. 2.3.4 in connection with radiative $\Upsilon$ decays. The fragmentation contribution becomes more important at low values of the photon transverse momentum [89].

A complication in comparing theory and experiment is that an isolation cut is necessary experimentally in order to separate prompt photons from those due to hadron decays. A typical cut [84] requires less than 2 GeV of hadronic energy in a cone around the photon of size $R = 0.7$, defined as in eq. (46). This type of cut affects both the direct and fragmentation contributions to the photon spectrum.

The CDF and D0 data, shown in figure 22 [84], agree with next-to-leading order predictions, except possibly at lower transverse momenta ($p_T^\gamma <$ 30 GeV), where a steeper slope and a consequent excess of photons are seen. Figure 23 shows a comparison with the calculation of Glück et al. [89], in which the relative discrepancy is shown after dividing by the theoretical prediction. These authors find that, using their parametrizations of the fragmentation functions [88] and parton distributions [90], the disagreement at low $p_T^\gamma$ is reduced when the fragmentation contribution is included. More accurate data below 20 GeV will be needed to decide whether the slope is now given correctly. This type of comparison should also be performed with other parton distributions, for example those of MRS [51], so see whether the improved agreement is a special feature of the GRV distributions [90].

### *5.3. Diffractive jet production*

The data from HERA on deep inelastic events containing a large rapidity gap between the proton remnant and the current jet suggest that about 10% of the cross section is diffractive [11, 91]. The term "diffractive" is used here simply to mean that the process has many of the features of elastic scattering: exchange of vacuum quantum numbers, flat energy dependence and sharp peaking at small momentum transfers. The object exchanged in diffractive processes is called the pomeron. Thus it may be possible to interpret the rapidity gap events at HERA as deep inelastic scattering on a pomeron.

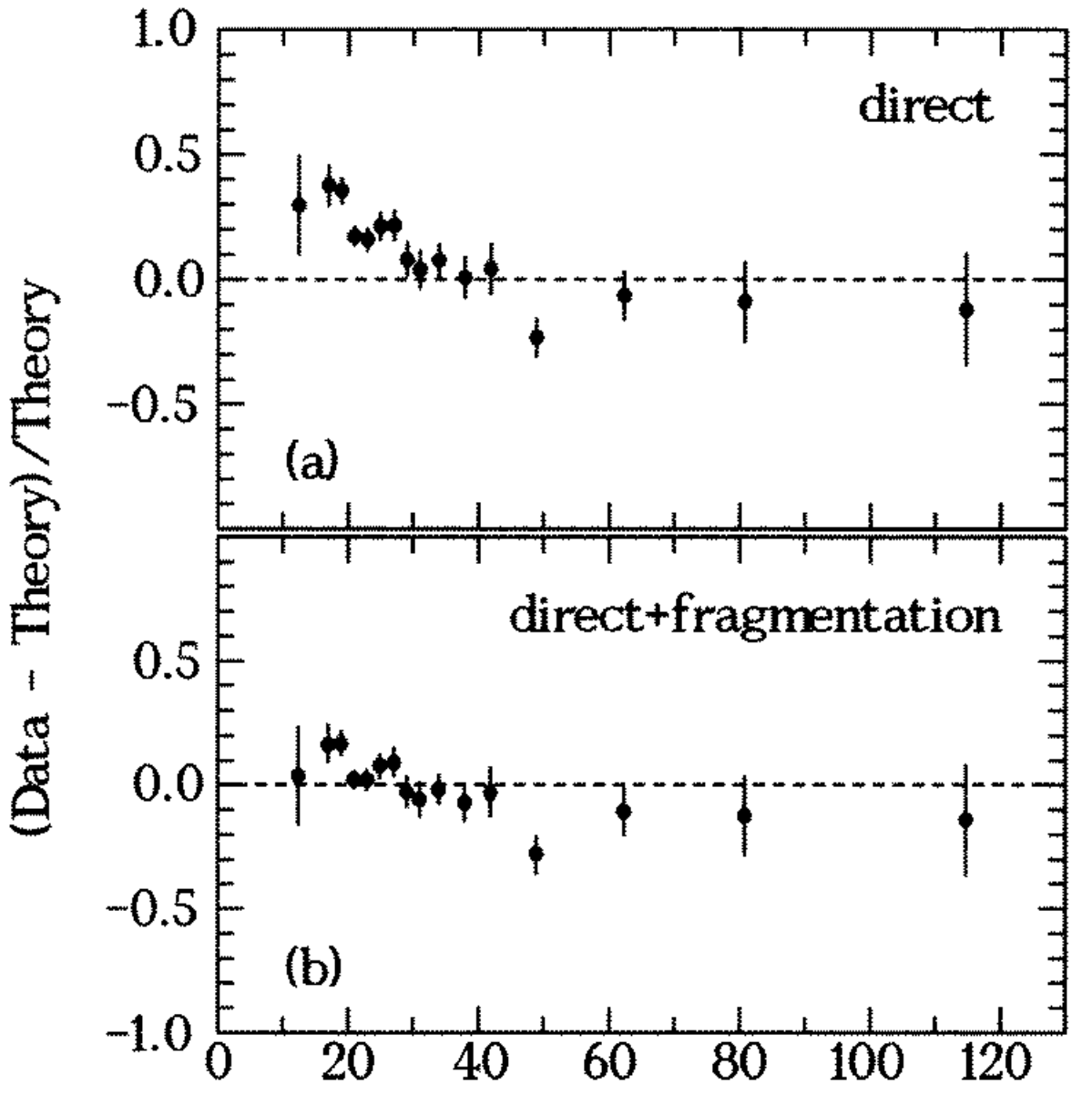

**Figure 23.** Comparisons between experiment and theoretical predictions for prompt photon production. The abscissa is photon transverse momentum in GeV/c.

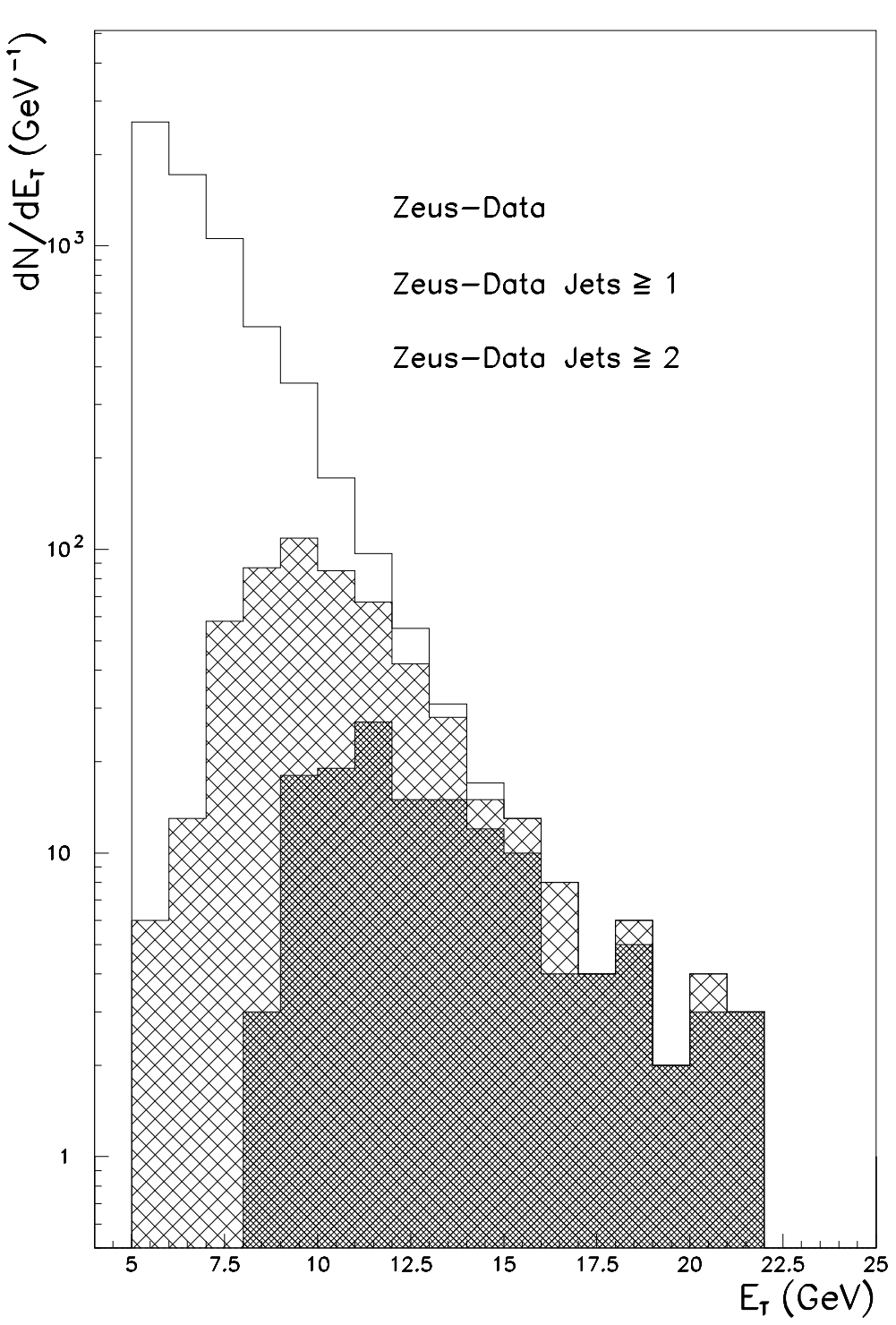

**Figure 24.** Jet fractions in photoproduction events with $\eta_{\rm max} \leq 1.5$, as a function of transverse energy.

As seen first in $p\bar{p}$ collisions by the UA8 collaboration [92], and confirmed now by the HERA experiments [11, 93, 94], jet production is possible in diffractive processes. This is most naturally interpreted as evidence of pointlike substructure in the pomeron. What is found at HERA is that the cross section for rapidity gap events (those with a maximum observed hadron rapidity $\eta_{\rm max} \leq 1.5$) is dominated at large transverse energy by production of at least two jets (figure 24). Note that the data shown are actually for photoproduction [93]; those for deep inelastic scattering are similar in this respect [94] but the statistics are lower. As soon as a large transverse energy is demanded the process enters the hard scattering regime, irrespective of the value of $Q^2$, so the pointlike substructure of the pomeron can be equally well explored in photoproduction.

We can expect this new source of jets to be a rich field of future investigation. Because the nature of the pomeron is still quite mysterious, the properties of the produced jets needs to be compared carefully with those in ordinary non-diffractive photoproduction and deep inelastic scattering. This applies not only to the jets with large transverse momenta but, if possible, also (especially) to the "pomeron remnant" jet.

## 6. Conclusions

My general conclusion from the wide range of comparisons reviewed here is that QCD is in good shape as the definitive theory of strong interactions. Its coupling constant $\alpha_S$ cleary displays the running behaviour that is necessary for asymptotic freedom, and a reasonably consistent value, summarized in eq. (26) and figure 6, emerges from a great variety of perturbative and (in the lattice QCD approach) non-perturbative phenomena.

The colour structure of QCD is now also well confirmed by studies of 2-, 3- and 4-jet final states at LEP, as summarized by figure 21. Although the simple asymptotic relationship between the colour factors and the fragmentation properties of quark and gluon jets is far from being realized at present energies, these properties do show clearly the expected qualitative differences. Quantitative agreement is also achieved in quark/gluon jet comparisons for properties insensitive to hadronization, such as sub-jet multiplicities at sufficiently coarse resolution.

Jets in $e^+e^-$ final states have been studied in great detail and the main features of event shape distributions, jet rates, jet profiles, etc. are now well understood. Similar studies in $p\bar{p}$ collisions are still in progress, and in ep collisions at HERA they are just beginning. We can expect many interesting phenomena in ep jet physics, for example a wealth of new information on jet production in diffractive processes. In all these studies, QCD Monte Carlo programs have played and will continue to play a valuable rôle, incorporating ever more sophisticated approximations [95] to the true predictions of the theory.

While bearing all these successes in mind, we should not ignore the problems that QCD still faces. Amongst the possible difficulties, we have discussed the slight systematic discrepancies between $\alpha_S$ values obtained from different classes of measurements. Another mild problem is the apparent underestimation of heavy quark production cross sections, most notably that of the top quark [96].

The greatest problem, however, remains our poor understanding of non-perturbative processes in QCD. At present only the lattice approach can provide reliable quantitative predictions outside the realm of perturbation theory, and lattice calculations are laborious, costly and limited in scope. The outstanding non-perturbative problem of jet physics is to understand the hadronization process, which seems well beyond the present capabilities of lattice QCD. As long as we have to rely on hadronization models, our quantitative understanding of jets cannot progress much beyond the current level. At least we need a framework for parametrizing power corrections due to hadronization, like the operator product expansion for higher twist corrections in deep inelastic scattering. In this connection, studies of the divergence of the perturbative expansion at very high orders might provide some clues to the non-perturbative behaviour of the theory.

## Apologies

Regrettably, limitations of space and time have prevented me from discussing or even referring to many interesting contributions relevant to QCD and jet physics, only some of which were covered by parallel session talks. Areas of strong activity not already mentioned include: $\gamma\gamma$ jet physics, prompt photons in $e^+e^-$ final states, tests of flavour independence of $\alpha_S$, studies of rapidity gaps and coherence effects [97], heavy quark [96, 98, 99] and Higgs boson production [100], spin physics and jet handedness [101], single-particle yields and spectra [102], Bose-Einstein correlations [103], intermittency [104], and connections between QCD and solvable models [105].

## Acknowledgements

It is a pleasure to thank many people for their help and valuable comments, especially S. Bethke, G. Cowan, C.T.H. Davies, R.K. Ellis, V. Jamieson, I.G. Knowles and D.R. Ward.